\documentclass[12pt]{article}

\usepackage{amsmath,amssymb}
\usepackage{graphicx,a4,psfrag,here,epsfig,epsf}
\usepackage{pstricks}
\usepackage{pst-node}
\usepackage{citesort}
\allowdisplaybreaks[1]

\begin{document}
%%%%%%%%%%%%%%%%%%%%%%%%%%%%%%%%%%%%%%%%%%%%%%%%%%%%%%%%%%%%%%%%%%%%%%%%%%%%%
\newcommand{\sh}{\not\!}
\newcommand{\lag}{{\mathcal L}}
\newcommand{\ol}{\overline}
\newcommand{\co}{\; \; ,}
\newcommand{\per}{ \; .}
\newcommand{\nn}{\nonumber}

\newcommand{\lt}{\left}% NEW
\newcommand{\rt}{\right}
\renewcommand{\d}{\partial}
\newcommand{\fc}{\frac}

\newcommand{\rR}{R}
\newcommand{\rr}{{\cal R}}
\newcommand{\mc}{\multicolumn{2}{|c|}}
\newcommand{\mcc}{\multicolumn{3}{|c|}}
\newcommand{\nl}{\!\! & \!\!}
\newcommand{\av}[1]{\langle {#1}\rangle}

\def\beq{\begin{equation}}
\def\eeq{\end{equation}}
\def\bea{\begin{eqnarray}}
\def\eea{\end{eqnarray}}
\def\eq{\begin{eqnarray}}
\def\en{\end{eqnarray}}
\newcommand{\bes}{\begin{split}}
\newcommand{\ees}{\end{split}}
\newcommand{\bed}{\begin{displaymath}}
\newcommand{\eed}{\end{displaymath}}
\def\prop{{\mathcal D}}

\newcommand{\mpi}{M_\pi^2}
\newcommand{\mk}{M_K^2}
\newcommand{\me}{M_\eta^2}
\newcommand{\equ}{\,=\,}
\newcommand{\no}{\nonumber}
\newcommand{\noh}{\nonumber \hskip -10cm}
\newcommand{\Order}{\mathcal{O}}
\newcommand{\Lagr}{\mathcal{L}}
\newcommand{\M}{\mathcal{M}}
\numberwithin{equation}{section}
\newcommand{\ed}{\end{document}}
\newcommand{\scs}{\, , \,}
\newcommand{\sem}{\, ; \,}
\newcommand{\nnnl}{\nonumber\\}
\newcommand{\fs}{\, . \,}
\newcommand{\be}{\begin{eqnarray}}
\newcommand{\ee}{\end{eqnarray}}
\newcommand{\bra}{\right\rangle}
\newcommand{\bla}{\left\langle}
\newcommand{\word}[1]{{\mbox{{#1}\,}}}
\newcommand{\words}[1]{{\mbox{\small{#1}}}}
\newcommand{\wordt}[1]{{\mbox{\tiny{#1}}}}

\def\query#1{\marginpar{\begin{flushleft}\footnotesize#1\end{flushleft}}}%

\begin{titlepage}

\begin{flushright}
UW preprint INT-PUB-13-025
\end{flushright}

\vspace{1cm}
\begin{center}{\Large\bf Partial twisting for scalar mesons}

\vspace{0.5cm}
\today

\vspace{0.5cm}
Dimitri Agadjanov$^{a,b}$, Ulf-G. Mei{\ss}ner$^{a,c}$
and Akaki Rusetsky$^a$

\vspace{2em}
\footnotesize{\begin{tabular}{c}
$^a\,$ Helmholtz-Institut f\"ur Strahlen- und Kernphysik (Theorie) and\\
Bethe Center for Theoretical Physics,\\
 $\hspace{2mm}$   Universit\"at Bonn,
Nussallee 14-16, D-53115 Bonn, Germany\\[2mm]
$^b\,$ St. Andrew the First-Called Georgian University of the Patriarchate of Georgia,\\ Chavchavadze Ave. 53a, 0162, Tbilisi, Georgia\\[2mm]
$^c\,$ 
Institute for Advanced Simulation (IAS-4), Institut f\"ur Kernphysik 
(IKP-3) and\\ J\"ulich Center for Hadron Physics,
Forschungszentrum J\"ulich, D-52425 J\"ulich, Germany
\end{tabular}  }

\vspace{1cm}

\begin{abstract}

\noindent
The possibility of imposing partially twisted boundary 
conditions is investigated for the scalar sector of lattice QCD.
According to the commonly shared belief, the presence of  
quark-antiquark annihilation diagrams in the intermediate state 
generally 
hinders the use of the partial twisting. Using effective field
theory techniques in a finite volume, and studying the scalar sector
of QCD with total isospin $I=1$,
we however demonstrate that 
partial twisting can still be performed, despite the fact that 
annihilation diagrams
are present. The reason for this are delicate cancellations, which
emerge due to the graded symmetry in partially quenched QCD with 
valence, sea and ghost quarks. The modified L\"uscher equation in case 
of partial twisting is given.
\end{abstract}

\vspace{1cm}
\footnotesize{\begin{tabular}{ll}
{\bf{Pacs:}}$\!\!\!\!$& 12.38.Gc, 12.39.Fe, 13.75.Lb
\\
{\bf{Keywords:}}$\!\!\!\!$& Lattice QCD, Partially Quenched Chiral
Perturbation Theory,\\ &Non-relativistic EFT,
L\"uscher equation, scalar mesons
\\
\end{tabular}}
\end{center}
\end{titlepage}

\setcounter{page}{2}

%\tableofcontents

\section{Introduction}
\label{sec:intro}

Investigating the scalar sector of QCD in the region below and around 1~GeV 
on the lattice enables one to gain important information about the low-energy
behavior of strong interactions. A few groups have addressed this 
problem in the recent years (see, e.g.,~\cite{Kunihiro,Hart,Prelovsek,Urbach}).
Note that  carrying out simulations in the scalar sector is a very challenging task
by itself as many of these states share the quantum numbers of the vacuum. 
In addition, it is known that the particles, whose properties are investigated in
these simulations, are resonances. Consequently, in order to perform 
the extraction of their mass and width from the data, one has to 
apply the L\"uscher approach~\cite{Luescher-torus} that implies carrying out simulations at different
volumes, complicating further an already difficult problem. Moreover, 
in case of the $f_0(980)$ and $a_0(980)$ mesons, the analysis has to be done
by using a {\em coupled-channel} L\"uscher equation~\cite{Lage-scalar,Oset-scalar1,Oset-scalar2}, which includes $\pi\pi/K\bar K$ and $\pi\eta/K\bar K$ channels
for  total isospin $I=0$ and $I=1$, respectively. The resonances are 
very close to the $K\bar K$ (inelastic) threshold, which has the unpleasant
property of ``masking'' the avoided level crossing that serves as a signature
of the presence of a resonance in a finite 
volume~\cite{Lage-scalar,Oset-scalar1,Oset-scalar2}. 

Here, one should also mention that the mass and width are not the only
quantities one is interested in case of scalar resonances. The nature of 
these states is not well established in phenomenology and is being debated at present, with the
arguments given in favor of their interpretation as tetraquark states
(see, e.g.,~\cite{Jaffe:1976ig,Black:1998wt,Achasov:2002ir,Pelaez:2004xp}),  
as $K\bar K$ molecules~\cite{Weinstein:1982gc,Oller:1997ng,oop}, or 
as a combination of a bare pole and the rescattering 
contribution~\cite{Oller:1998zr} (see also Refs.~\cite{sumrules,Janssen:1994wn}
for more information on this issue). In view of the conflicting interpretations,
it is interesting to study the signatures of a possible exotic behavior,
e.g., applying Weinberg's compositeness condition or the pole counting criterion
(see, e.g.,~\cite{Weinberg:1963zz,Morgan:1992ge,Tornqvist:1994ji,Morgan:1993td,Baru:2003qq,Baru:2004xg,Hanhart:2006nr}), or investigating the quark mass dependence of the resonance pole position~\cite{Lage-scalar}. It is possible to ``translate'' all these criteria into the language of lattice QCD. However, testing them in the real simulations would require much more
data at different volumes and at a much higher precision than it is at our disposal at present.  

Summarizing all the facts above, it is legitimate to ask, whether -- 
given our present capabilities -- the 
extraction of the properties of scalar resonances on the lattice 
can be realistically done with a sufficient rigor and yield clean and unambiguous results in the nearest future.

In Refs.~\cite{Lage-scalar,Oset-scalar1,Oset-scalar2} it has been pointed
out that using twisted boundary conditions in lattice simulations~\cite{twisted,Sachrajda,Chen} can provide an important advantage in the scalar meson sector.
First and foremost, varying the twisting angle $\mbox{\boldmath $\theta$}$ can 
substitute for 
simulations at different volumes and provide data of energy levels, 
which should be fitted in order to determine the resonance pole position. Note
that the same effect can be achieved by carrying out simulations at a non-zero
total momentum. However, whereas the components of the lattice momentum are
given by integer numbers in the units of $2\pi/L$, where $L$ is the size of the
finite box, the twisting angle can be varied continuously. Another advantage
is provided by the fact that twisting allows one to effectively move the 
threshold away from the resonance pole location. In order to illustrate this,
consider an example when the $s$-quark is twisted in the simulations, whereas $u$ and $d$ quarks still obey periodic boundary conditions~\cite{Lage-scalar,Oset-scalar1,Oset-scalar2}. Assume, in addition, that the system is in the center-of-mass (CM) frame. In this example, the $K$ and $\bar K$ mesons in the $K\bar K$
intermediate state acquire 3-momenta, opposite in direction and having equal magnitude, proportional to 
  $|\mbox{\boldmath $\theta$}|$. Hence, the energy of the ground state of the $K\bar K$ pair goes up, whereas the resonance, 
which corresponds to a true pole in the $S$-matrix, stays, by
definition, at the same position. 
For the volumes, which are currently used in lattice simulations, the upward displacement of the
$K\bar K$ threshold would be a large effect. Consequently, it could be expected that, fitting  twisted lattice data, one would achieve a more accurate extraction
of the resonance pole position than in the case of periodic boundary conditions, when the threshold and the resonance are very close. Note that this conjecture has been fully confirmed in Refs.~\cite{Oset-scalar1,Oset-scalar2} by performing fits to ``synthetic'' data sets.

\begin{figure}[t]
\begin{center}
\includegraphics[width=6.cm]{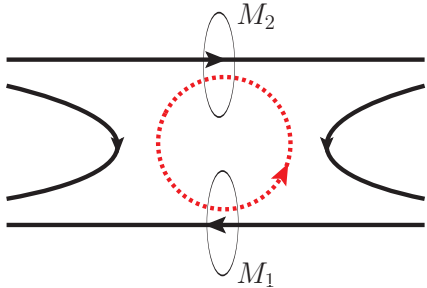}
\end{center}
\caption{An example of an annihilation diagram in meson-meson scattering.
The full and dashed lines denote valence and sea quarks, respectively.
The intermediate state for this diagram consists of two mesons $M_1$ and $M_2$
with one valence and one sea quark.}
\label{fig:annihilation}
\end{figure}

There is, however, an important caveat in the arguments above. Imposing 
twisted boundary conditions in lattice simulations implies the calculation
of gauge configurations anew. This is a very expensive enterprise.
The majority of simulations up to day are done by applying the so-called
partial twisting, i.e., twisting only the valence quarks and leaving 
the configurations the same. It can be proven (see, e.g.~\cite{Sachrajda,Chen})
that in many cases the results obtained by using partial and full twisting
coincide up to exponentially suppressed terms. This happens when
there are no annihilation diagrams, i.e., the diagrams where the {\em valence}
quark-antiquark pair from the initial state can annihilate and a pair of
the {\em sea} quark-antiquark is produced, which obey a different boundary
condition (see Fig.~\ref{fig:annihilation}). However, it is easy to verify 
that, in case of scalar mesons,
the annihilation diagrams do appear. Consequently, following the arguments of
Refs.~\cite{Sachrajda,Chen}, one had to conclude that the partial twisting
in this case is useless -- one has either to perform a full twisting, or to give
it up.

We consider this conclusion premature. One could look at the problem from
a different point of view. It is definitely not possible to prove in general
that in this case the partial and full twisting lead to the same result.
Could one find a {\em modified} L\"uscher equation, which corresponds to the
case of partial twisting? Does this equation enable one to still extract the
physically interesting information about the scattering $S$-matrix elements 
in the finite volume? If the answer to this question is yes, using partial
twisting in lattice simulations can be justified.

In this paper we do not give a full-fledged solution of the problem. Rather,
we have chosen to concentrate on one particular example, namely, the $a_0(980)$,
which is an $S$-wave resonance with the isospin $I=1$, and solve this problem
to the end. Possible mixing to other partial waves is neglected.
The inclusion of higher partial waves
forms a subject of a separate investigation which will be carried out in 
the future.

A brief outline of the method is as follows. It is well known that L\"uscher's
equation can be most easily derived by using non-relativistic EFT framework
in a finite volume~\cite{Beane,Lage-piN,Hoja}. Twisting at the quark level
can be straightforwardly implemented at the hadronic level: the hadrons acquire
additional momenta, proportional to the twisting angle $\mbox{\boldmath $\theta$}$.
The expression for the zeta-function in the L\"uscher equation also changes
in a well-defined way,
whereas the non-relativistic potentials, which encode the short-range dynamics,
are $\mbox{\boldmath $\theta$}$-independent. All this gives the 
L\"uscher equation in case of twisted boundary conditions.

The case of partially twisted boundary conditions can be considered analogously.
The spectrum of the effective theory now contains much more hadrons, consisting of valence, sea and ghost quarks (see, e.g.,~\cite{Sharpe}). Boundary conditions for each hadron are determined by the boundary conditions on its constituents,
so the $\mbox{\boldmath $\theta$}$-dependence of the zeta-functions, entering
the L\"uscher equation, is uniquely defined also in this case. The crucial
observation, which enables one to arrive at a tractable form of the L\"uscher
equation, is that the symmetries, which are present in the theory in the
infinite volume, relate the potentials in valence, sea and ghost sectors
(the masses of valence, sea and ghost quarks are taken equal). It can be shown
that the L\"uscher equation can be reduced to the one that contains the 
potentials only in the physical (valence) sector and can thus be used to 
analyze the lattice data.

The layout of the paper is as follows. In section~\ref{sec:framework}
we describe the effective field theory (EFT) framework for partially twisted
QCD -- first, in the infinite volume. In section~\ref{sec:symmetries}
we discuss in detail the constraints imposed by the symmetries on the matrix
elements of the effective non-relativistic potential. In doing this, we
first neglect the neutral meson mixing beyond tree level. 
In section~\ref{sec:luescher}
the L\"uscher equation in case of the partially twisted boundary conditions
is derived. Possible applications in the simulations in the scalar sector
are discussed. In section~\ref{sec:mixing} we clear the remaining loopholes
by discussing the mixing to all orders in this framework and 
show that the results are not affected. Finally, 
section~\ref{sec:concl} contains our conclusions and outlook.

\section{The effective field theory framework}
\label{sec:framework}
 
In order to obtain the spectrum, one usually studies the behavior of certain
correlators at a large Euclidean time separation $t$:
\eq\label{eq:correlator}
C(t)=\langle {\cal O}(t){\cal O}^\dagger(0)\rangle
=\frac{1}{Z}\int {\cal D}U{\cal D}\psi{\cal D}\bar\psi\,
 {\cal O}(t){\cal O}^\dagger(0)\,\exp\biggl\{
-S_G-\int d^4x\bar\psi(\not\!\! D+m)\psi\biggr\}\, ,
\en
where $S_G$ stands for the gluon action functional, and ${\cal O}(t),
{\cal O}^\dagger(t)$ are
appropriate source/sink operators, which have a non-zero overlap with the 
physical states of interest. At this stage, we do not specify the explicit
form of these operators -- these can be, for example, quark-antiquark or
two meson operators, etc.

In order to distinguish between valence and sea quarks, we use the standard
 trick
(see, e.g.,~\cite{Sharpe} and references therein), rewriting the 
above path integral in the following manner
\eq\label{eq:trick}
\hspace*{-.5cm}C(t)&\!\!\!=\!\!\!&\frac{1}{Z}\int {\cal D}U{\cal D}\psi_{\sf v}{\cal D}\bar\psi_{\sf v}{\cal D}\psi_{\sf s}{\cal D}\bar\psi_{\sf s}{\cal D}\psi_{\sf g}{\cal D}\bar\psi_{\sf g}\,
 {\cal O}_{\sf v}(t){\cal O}_{\sf v}^\dagger(0)
\nonumber\\[2mm]
\hspace*{-.5cm}&\!\!\!\times\!\!\!&\exp\biggl\{
-S_G-\int d^4x\biggl[\bar\psi_{\sf v}(\not\!\! D+m_{\sf val})\psi_{\sf v}+\bar\psi_{\sf s}(\not\!\! D+m_{\sf sea})\psi_{\sf s}
+\bar\psi_{\sf g}(\not\!\! D+m_{\sf gh})\psi_{\sf g}\biggr]\biggr\}\, .
\en
Here, the subscripts ``v,'' ``s'' and ``g'' stand for valence, sea and ghost
quarks, the latter being described by {\em commuting} spinor fields. 
After performing the path integral over quarks, it is seen that the fermion
determinant, coming from valence quarks, is exactly cancelled by the one
from the ghost quarks, and the expression, given in Eq.~(\ref{eq:correlator}),
is reproduced.

In order to describe the situation with partially twisted boundary conditions,
one imposes twisted boundary conditions on the valence and ghost quarks and
periodic boundary conditions on the sea quarks. The masses of all species of
quarks are taken equal, in difference to the partially quenched case where
$m=m_{\sf val}=m_{\sf gh}\neq m_{\sf sea}$. Note that
$m_{\sf val},m_{\sf gh},m_{\sf sea}$ are matrices in flavor
space. Note also that we assume isospin symmetry
throughout the paper $m_u=m_d=\hat m\neq m_s$. 

In the chiral limit, the infinite-volume theory is invariant under the
 graded symmetry group $SU(2N|N)_L\times SU(2N|N)_R\times U(1)_V$, where $N=3$ is the number
of light flavors.
The low-energy effective Lagrangian, corresponding to the case of 
partially twisted boundary conditions, contains the matrix 
$U=\exp\{i\sqrt{2}\Phi/F\}$ of the pseudo-Goldstone fields $\Phi$, 
which transforms under this group as
\eq
U\to LUR^\dagger\, ,\quad\quad L,R\in SU(2N|N)\, .
\en
The Hermitian matrix $\Phi$ has the following representation
\eq
\Phi=
\begin{pmatrix}
M_{\sf vv} & M_{\sf sv}^\dagger & M_{\sf gv}^\dagger \\ 
M_{\sf sv} & M_{\sf ss} & M_{\sf gs}^\dagger \\ 
M_{\sf gv} & M_{\sf gs} & M_{\sf gg}
\end{pmatrix}\, .
\en 
Here, each of the entries is itself a $N\times N$ matrix in flavor 
space, containing meson fields built up from certain quark species (e.g., from valence quark and valence antiquark, from sea quark and ghost antiquark, and so on).
The fields $M_{\sf gv}$ and $M_{\sf gs}$ are anti-commuting pseudoscalar fields
(ghost mesons). Further, the matrix $\Phi$ obeys the condition~\cite{Sharpe}
\eq\label{eq:str}
\mbox{str}\,\Phi=\mbox{tr}\,(M_{\sf vv}+M_{\sf ss}-M_{\sf gg})=0\, ,
\en
where ``str'' stands for the supertrace.

The effective chiral Lagrangian takes the form
\eq\label{eq:L}
{\cal L}=\frac{F_0^2}{4}\mbox{str}\,(\partial_\mu U\partial^\mu U^\dagger)
-\frac{F_0^2}{4}\,\mbox{str}(\chi U+U\chi^\dagger)+\mbox{higher-order terms,}
\en
where $\chi=2mB_0$ is proportional to the quark mass matrix.

In the infinite volume, the above theory is completely equivalent to ordinary
Chiral Perturbation Theory (ChPT), since the masses of the quarks of all species
are set equal. In a finite volume, the difference arises due to the different
boundary conditions, set on the different meson fields. These boundary 
conditions are uniquely determined by the boundary conditions imposed on the
constituents.

We do not intend to use the framework of the partially twisted ChPT to
carry out explicit calculations. We need this framework only to facilitate
 the derivation of the L\"uscher equation. To this end, let us consider 
large boxes with $L\gg M_\pi^{-1}$, where $M_\pi$ is the lightest mass in the system (the pion mass). The characteristic 3-momenta in such a box
 are much smaller than all masses -- consequently, the system can be described
by a non-relativistic EFT, whose low-energy couplings are
consistently matched to the relativistic theory with the Lagrangian
given in Eq.~(\ref{eq:L}) (for a detailed review of
the non-relativistic theory in the infinite volume, we refer the reader, e.g., to the Refs.~\cite{physrep,cusps}; non-relativistic effective field theories in a finite volume are considered in Refs.~\cite{Beane,Lage-piN,Hoja}.).
The two-body scattering $T$-matrix in the non-relativistic theory obeys the
multi-channel Lippmann-Schwinger (LS) equation (for simplicity, we write down this equation in the CM frame)
\eq\label{eq:LS}
T_{\alpha\beta}({\bf p},{\bf q};P_0)=V_{\alpha\beta}({\bf p},{\bf q})+
\sum_\gamma\int\frac{d^d{\bf k}}{(2\pi)^d}\,
\frac{V_{\alpha\gamma}({\bf p},{\bf q})T_{\gamma\beta}({\bf p},{\bf q};P_0)}
{2w_1^{(\gamma)}({\bf k})2w_2^{(\gamma)}({\bf k})(w_1^{(\gamma)}({\bf k})+w_2^{(\gamma)}({\bf k})
-P_0-i0)}\, ,
\en
where the sum runs over all two-body channels labeled by the index $\gamma$, and 
$w_1^{(\gamma)}({\bf k})$, $w_2^{(\gamma)}({\bf k})$ stand for the (relativistic) energies of the first
and the second particle in this channel. The potentials 
$V_{\alpha\beta}({\bf p},{\bf q})$ encode the short-range dynamics, including inelastic
many-body channels, which are closed at low energies\footnote{There is a caveat
in this argument. For example, there are multi-pion channels
 below $K\bar K$ channel. However, since the couplings to these channels are very weak, they can be safely ignored without changing the result. For more discussion on this issue, see Ref.~\cite{Lage-scalar}.}.
These potentials are constructed perturbatively and contain the couplings of the non-relativistic effective Lagrangian.

We use dimensional regularization throughout. In this regularization, the
potentials $V_{\alpha\beta}({\bf p},{\bf q})$ coincide with the $K$-matrix elements.
Expanding into the partial waves gives:
\eq
V_{\alpha\beta}({\bf p},{\bf q})=4\pi\sum_{lm}
Y_{lm}(\hat{\bf p})V_{\alpha\beta}^l(|{\bf p}|,|{\bf q}|)Y^*_{lm}(\hat {\bf q})\, ,
\en
where $\hat {\bf k}$ denotes a unit vector in the direction of ${\bf k}$.
It is easy to see that in the elastic region for the channel $\alpha$,
on the energy shell where $|{\bf p}|=|{\bf q}|=q_0$,
\eq\label{eq:tan}
V_{\alpha\alpha}^l(q_0,q_0)=\frac{8\pi P_0}{q_0}\,\tan\delta_l^{(\alpha)}(q_0)\, ,
\quad\quad P_0=w_1^{(\alpha)}(q_0)+w_2^{(\alpha)}(q_0)\, ,
\en
where $\delta_l^{(\alpha)}$ denotes the elastic scattering phase shift with angular
momentum $l$. In the following, we shall neglect all partial waves except $l=0$.
The inclusion of partial-wave mixing will be considered in the future.  

When the relativistic theory, described by the Lagrangian given in
Eq.~(\ref{eq:L}), is matched to the non-relativistic EFT, a complication
arises, which stems from the mixing of the states containing neutral mesons. 
Namely, in the equation~(\ref{eq:LS}), the states $\alpha,\beta,\gamma$ correspond to the 
{\em physical} two-particle states. These are not always described
by the
meson fields which are present in the matrix $\Phi$. The reason for this
is that not all the components of $\Phi$ are independent due to the condition
$\mbox{str}\,\Phi=0$.

In order to study the issue of mixing, let us again start with the relativistic
theory described by the Lagrangian in Eq.~(\ref{eq:L}). We restrict ourselves
first to order $p^2$, and retain only diagonal terms in the
matrix $\Phi=\mbox{diag}\,(\phi_1,\ldots,\phi_9)$ 
(the non-diagonal terms do not mix). The quadratic piece in the $O(p^2)$ 
Lagrangian takes the form
\eq\label{eq:1-9}
{\cal L}^{(2)}_0&\!\!\!=\!\!\!&\frac{1}{2}\,
\sum_{i=1}^6(\partial_\mu\phi_i)^2
-\frac{1}{2}\,\sum_{i=7}^9(\partial_\mu\phi_i)^2
\nonumber\\[2mm]
&\!\!\!-\!\!\!&\frac{M^2}{2}\,(\phi_1^2+\phi_2^2+\phi_3^2+\phi_4^2-\phi_7^2-\phi_8^2)-\frac{M_s^2}{2}\,(\phi_3^2+\phi_6^2-\phi_9^2)\,
,\nonumber\\[2mm]
M^2&\!\!\!=\!\!\!&2\hat m B_0\, ,\quad\quad M_s^2=2m_sB_0\, .
\en
Introducing the following linear combinations
\eq\label{eq:phi-omega}
\phi_1&=&\frac{1}{\sqrt{2}}\,\omega_1-\frac{1}{\sqrt{6}}\,\omega_2
-\frac{1}{2}\,\omega_5+\frac{1}{2}\,\omega_8\, ,
\nonumber\\[2mm]
\phi_2&=&-\frac{1}{\sqrt{2}}\,\omega_1-\frac{1}{\sqrt{6}}\,\omega_2
-\frac{1}{2}\,\omega_5+\frac{1}{2}\,\omega_8\, ,
\nonumber\\[2mm]
\phi_3&=&\frac{\sqrt{6}}{3}\,\omega_2+\frac{1}{\sqrt{2}}\,\omega_3-\frac{1}{\sqrt{2}}\,\omega_6\, ,
\nonumber\\[2mm]
\phi_4&=&-\frac{1}{\sqrt{6}}\,\omega_2+\frac{1}{\sqrt{2}}\,\omega_4
+\frac{1}{2}\,\omega_5+\frac{1}{2}\,\omega_8\, ,
\nonumber\\[2mm]
\phi_5&=&-\frac{1}{\sqrt{6}}\,\omega_2-\frac{1}{\sqrt{2}}\,\omega_4
+\frac{1}{2}\,\omega_5+\frac{1}{2}\,\omega_8\, ,
\nonumber\\[2mm]
\phi_6&=&\frac{\sqrt{6}}{3}\omega_2+\frac{1}{\sqrt{2}}\,\omega_3
+\frac{1}{\sqrt{2}}\,\omega_6\, ,
\nonumber\\[2mm]
\phi_7&=&-\frac{1}{\sqrt{6}}\,\omega_2+\frac{1}{\sqrt{2}}\,\omega_7+\omega_8
\nonumber\\[2mm]
\phi_8&=&-\frac{1}{\sqrt{6}}\,\omega_2-\frac{1}{\sqrt{2}}\,\omega_7+\omega_8
\nonumber\\[2mm]
\phi_9&=&\frac{\sqrt{6}}{3}\,\omega_2+\sqrt{2}\,\omega_3\, ,
\en
it is straightforward to check that the quadratic piece of the
 Lagrangian can be rewritten in terms of the
fields $\omega_1,\ldots,\omega_8$:
\eq\label{eq:L-omega}
{\cal L}^{(2)}_0&\!\!\!=\!\!\!&\frac{1}{2}\,\biggl\{
(\partial_\mu\omega_1)^2
+(\partial_\mu\omega_2)^2
-(\partial_\mu\omega_3)^2
+(\partial_\mu\omega_4)^2
+(\partial_\mu\omega_5)^2
+(\partial_\mu\omega_6)^2
-(\partial_\mu\omega_7)^2
-(\partial_\mu\omega_8)^2\biggr\}
\nonumber\\[2mm]
&\!\!\!-\!\!\!&\frac{M^2}{2}\,(\omega_1^2+\omega_4^2+\omega_5^2-\omega_7^2-\omega_8^2)
+\frac{M_s^2}{2}\,(\omega_3^2-\omega_6^2)-\frac{M_\eta^2}{2}\, \omega_2^2\, ,
\en
where $M_\eta^2=\frac{2}{3}\,M_s^2+\frac{1}{3}\,M^2$. Note that the 
condition $\mbox{str}\,\Phi=0$ is automatically fulfilled for the fields
given by Eq.~(\ref{eq:phi-omega}). The eight fields $\omega_1,\ldots,\omega_8$
are unconstrained as opposed to the nine fields $\phi_1,\ldots,\phi_9$.
The propagators for the physical particles can be read off from 
Eq.~(\ref{eq:L-omega}). The fields $\omega_3,\omega_7,\omega_8$ are ghost
fields (they enter the Lagrangian with a ``wrong'' sign).

The fields $\omega_i$ describe physical particles at $O(p^2)$, and matching to
the non-relativistic theory is most easily performed in this basis. The
symmetry relations between various matrix elements get, however, very
complicated in this basis. In order to circumvent this problem, we have chosen
to work in another basis
\eq\label{eq:basis}
\pi^0_{\sf vv}&\!\!\!=\!\!\!&\omega_1\, ,\quad\quad
\pi^0_{\sf ss}=\omega_4\, ,\quad\quad
\pi^0_{\sf gg}=\omega_7\, ,
\nonumber\\[2mm]
\eta_{\sf vv}&\!\!\!=\!\!\!&-\omega_2+\frac{1}{\sqrt{6}}\,(-\omega_5+\omega_8-\sqrt{2}\omega_3
+\sqrt{2}\omega_6)\, ,\quad\quad
\eta'_{\sf vv}=\frac{1}{\sqrt{6}}\,(-\sqrt{2}\omega_5+\sqrt{2}\omega_8
+\omega_3-\omega_6)\, ,
\nonumber\\[2mm]
\eta_{\sf ss}&\!\!\!=\!\!\!&-\omega_2+\frac{1}{\sqrt{6}}\,(\omega_5+\omega_8-\sqrt{2}\omega_3
-\sqrt{2}\omega_6)\, ,\quad\quad
\eta'_{\sf ss}=\frac{1}{\sqrt{6}}\,(\sqrt{2}\omega_5+\sqrt{2}\omega_8
+\omega_3+\omega_6)\, ,
\nonumber\\[2mm]
\eta_{\sf gg}&\!\!\!=\!\!\!&-\omega_2+\frac{1}{\sqrt{6}}\,(2\omega_8-2\sqrt{2}\omega_3)\, ,
\quad\quad
\eta'_{\sf gg}=\frac{1}{\sqrt{3}}\,(2\omega_8
+\sqrt{2}\omega_3)\, .
\en
The propagator matrix is defined as
\eq
i\langle 0|T\phi_A(x)\phi_B(0)|0\rangle
=\int \frac{d^4p}{(2\pi)^4}\,e^{-ipx}\,D_{AB}(p)\, ,\quad\quad A,B={\sf vv,ss,gg}\, .
\en
Due to isospin symmetry this matrix is diagonal in the subspace with different species of $\pi^0$:
\eq\label{eq:pionprop}
D_{\pi^0_{\sf vv}\pi^0_{\sf vv}}(p)
=D_{\pi^0_{\sf ss}\pi^0_{\sf ss}}(p)
=-D_{\pi^0_{\sf gg}\pi^0_{\sf gg}}(p)=D_\pi\, ,
\en
However, different species of the $\eta$ and $\eta'$ mix. Defining the $2\times 2$ matrix
\eq\label{eq:etamatrix}
\Omega_{AB}(p)=\begin{pmatrix}
D_{\eta_A\eta_B}(p) &D_{\eta_A\eta'_B}(p) \\
D_{\eta'_A\eta_B}(p) &D_{\eta'_A\eta'_B}(p) 
\end{pmatrix}\, ,
\en
we get
\eq\label{eq:etaprop}
\Omega_{\sf vv,vv}&\!\!=\!\!&\Omega_{\sf ss,ss}=A\, ,\quad\quad
\Omega_{\sf gg,gg}=A+2X\, ,
\nonumber\\[2mm]
\Omega_{\sf vv,ss}&\!\!=\!\!&\Omega_{\sf vv,gg}=\Omega_{\sf ss,vv}=\Omega_{\sf ss,gg}=\Omega_{\sf gg,vv}=
\Omega_{\sf gg,ss}=A+X\, ,
\en
where
\eq\label{eq:AX}
A=\begin{pmatrix}
D_\eta & 0\\
0 & 0 
\end{pmatrix}\, ,\quad\quad
X=\begin{pmatrix}
-\frac{1}{3}\,D_\pi-\frac{2}{3}\,D_s & -\frac{\sqrt{2}}{3}\,(D_\pi-D_s) \\
-\frac{\sqrt{2}}{3}\,(D_\pi-D_s) &-\frac{2}{3}\,D_\pi-\frac{1}{3}\,D_s
\end{pmatrix}\, .
\en
In the above equations, the following notations were used:
\eq
D_\pi=\frac{1}{M^2-p^2}\, ,\quad\quad
D_\eta=\frac{1}{M_\eta^2-p^2}\, ,\quad\quad
D_s=\frac{1}{M_s^2-p^2}\, .
\en
If matching to the non-relativistic theory is performed in this basis, the
free two-particle Green function is no more diagonal in the channel basis
and the equation~(\ref{eq:LS}) is replaced by\footnote{We shall use Greek 
indices $\alpha,\beta,\gamma,\ldots$, to label channels in the basis where the
matrix of the two-point functions of the meson fields is diagonal.
This corresponds to working with the fields $\omega_1,\cdots,\omega_8$. On
the other hand, in the transformed basis (see Eq.~(\ref{eq:basis})), we label the channels by Latin letters $i,j,n,m,\ldots$. } 
\eq\label{eq:LS1}
T_{ij}({\bf p},{\bf q};P_0)=V_{ij}({\bf p},{\bf q})+
\sum_{nm}\int\frac{d^d{\bf k}}{(2\pi)^d}\,
V_{in}({\bf p},{\bf q})G_{nm}({\bf k};P_0)T_{mj}({\bf p},{\bf q};P_0)
\, .
\en
The entries of the matrix $G_{nm}$ can be easily determined by using
Eqs.~(\ref{eq:pionprop}), (\ref{eq:etamatrix}), (\ref{eq:etaprop}) and
(\ref{eq:AX}), see below. As already mentioned, the advantage of such a choice of
the basis is that the symmetry relations for the matrix elements
 $V_{ij}$, $T_{ij}$ are less complicated in this basis.

An important remark is in order. Up to now, we have considered the
mixing of the neutral states only at $O(p^2)$ in ChPT. 
The coefficients, e.g., in
Eq.~(\ref{eq:phi-omega}) will change, if higher-order terms are included.
How will this affect our expressions?

In order not to obscure the crucial physical arguments, we shall neglect
higher-order corrections for now.
At the end, we return to this question and show that the
final result remains unaffected by these corrections.

\section{Symmetries of the potential}
\label{sec:symmetries}

As already mentioned in the introduction, we concentrate on S-wave 
scattering in the sector with total isospin $I=1$. It is convenient to choose
$I_3=1$. Tracking the quarks of different species, flowing through the diagrams
describing meson-meson scattering, and starting from the state that contains only valence quarks, it is easy to see that 
the LS equation couples 11 different channels, as given in 
Table~1.

\renewcommand{\arraystretch}{1.8}
\setcounter{table}{0}
\begin{table}[t]\label{tab:channels}
\begin{eqnarray*}
\begin{array}{|l|l|l|}
\hline\hline
\mbox{Index} & \mbox{Channel} & \mbox{Quark content} \\
\hline
1 & |\pi^+_{\sf vv}\eta_{\sf vv}\rangle & 
-\frac{1}{\sqrt{6}}\,|(u_{\sf v}\bar d_{\sf v})
(u_{\sf v}\bar u_{\sf v}+d_{\sf v}\bar d_{\sf v}-2s_{\sf v}\bar s_{\sf v})\rangle\\
2 & |\pi^+_{\sf vv}\eta'_{\sf vv}\rangle & 
-\frac{1}{\sqrt{3}}\,|(u_{\sf v}\bar d_{\sf v})
(u_{\sf v}\bar u_{\sf v}+d_{\sf v}\bar d_{\sf v}+s_{\sf v}\bar s_{\sf v})\rangle\\
3 & |\pi^+_{\sf vv}\eta_{\sf ss}\rangle & 
-\frac{1}{\sqrt{6}}\,|(u_{\sf v}\bar d_{\sf v})
(u_{\sf s}\bar u_{\sf s}+d_{\sf s}\bar d_{\sf s}-2s_{\sf s}\bar s_{\sf s})\rangle\\
4 & |\pi^+_{\sf vv}\eta'_{\sf ss}\rangle & 
-\frac{1}{\sqrt{3}}\,|(u_{\sf v}\bar d_{\sf v})
(u_{\sf s}\bar u_{\sf s}+d_{\sf s}\bar d_{\sf s}+s_{\sf s}\bar s_{\sf s})\rangle\\
5 & |\pi^+_{\sf vv}\eta_{\sf gg}\rangle & 
-\frac{1}{\sqrt{6}}\,|(u_{\sf v}\bar d_{\sf v})
(u_{\sf g}\bar u_{\sf g}+d_{\sf g}\bar d_{\sf g}-2s_{\sf g}\bar s_{\sf g})\rangle\\
6 & |\pi^+_{\sf vv}\eta'_{\sf gg}\rangle & 
-\frac{1}{\sqrt{3}}\,|(u_{\sf v}\bar d_{\sf v})
(u_{\sf g}\bar u_{\sf g}+d_{\sf g}\bar d_{\sf g}+s_{\sf g}\bar s_{\sf g})\rangle\\
7 & |K^+_{\sf vv}\bar K^0_{\sf vv}\rangle & 
|(u_{\sf v}\bar s_{\sf v})(s_{\sf v}\bar d_{\sf v})\rangle\\
8 & |K^+_{\sf vs}\bar K^0_{\sf vs}\rangle & 
|(u_{\sf v}\bar s_{\sf s})(s_{\sf s}\bar d_{\sf v})\rangle\\
9 & |K^+_{\sf vg}\bar K^0_{\sf vg}\rangle & 
|(u_{\sf v}\bar s_{\sf g})(s_{\sf g}\bar d_{\sf v})\rangle\\
10 & |\pi^+_{\sf vs}\pi^0_{\sf vs}\rangle & 
\frac {1}{2}(-(u_{\sf v}\bar d_{\sf s})(u_{\sf s}\bar u_{\sf v}-d_{\sf s}\bar d_{\sf v})
+(u_{\sf v}\bar u_{\sf s}-d_{\sf v}\bar d_{\sf s})(u_{\sf s}\bar d_{\sf v})\rangle\\
11 & |\pi^+_{\sf vg}\pi^0_{\sf vg}\rangle & 
\frac {1}{2}(-(u_{\sf v}\bar d_{\sf g})(u_{\sf g}\bar u_{\sf v}-d_{\sf g}\bar d_{\sf v})
+(u_{\sf v}\bar u_{\sf g}-d_{\sf v}\bar d_{\sf g})(u_{\sf g}\bar d_{\sf v})\rangle\\
\hline\hline
\end{array}
\end{eqnarray*}
\caption{Scattering channels for the case of $I=I_3=1$.}
\end{table}

As immediately seen from this table, the valence sector couples to the sea and ghost sectors through the annihilation diagrams of the type shown in Fig.~\ref{fig:annihilation}. In addition, $\pi^+\pi^0$ states with quarks of different 
species are no more forbidden in the S-wave. Hence, in general, the partially
twisted L\"uscher equation will differ from the fully twisted one.

For comparison, let us consider a (trivial) example of meson scattering in the
channel with $I=2$, where the answer is already known. 
Take, for simplicity, $I_3=2$. 
In this case, starting in the valence quark sector,
one gets only one state $|\pi⁺_{\sf vv}\pi^+_{\sf vv}\rangle
=|(u_{\sf v}\bar d_{\sf v})(u_{\sf v}\bar d_{\sf v})\rangle$. Annihilation
diagrams are absent and, consequently, partial and full twisting are
equivalent up to exponentially suppressed contributions.
 
Using dimensional regularization, it is easy to see that the LS equation reduces
to an {\em algebraic} matrix equation (see, e.g.,
Refs.~\cite{Lage-piN,Hoja}). This equation relates the {\em on-shell} matrix
elements of $T$ and $V$.

The free Green function $G_{nm}$ in the channel with $I=1$ 
is given by the $11\times 11$ matrix
\eq\label{eq:G}
G=\begin{pmatrix}
B & B+Y & B+Y & O_1& O_1& O_1 & O_1& O_1\\
B+Y & B & B+Y & O_1& O_1& O_1&  O_1& O_1\\
B+Y & B+Y & B+2Y & O_1& O_1& O_1& O_1& O_1\\
O_1^T & O_1^T & O_1^T & K & 0 & 0 & 0 & 0 \\
O_1^T & O_1^T & O_1^T & 0 & K & 0 & 0 & 0 \\
O_1^T & O_1^T & O_1^T & 0 & 0 & -K & 0 & 0 \\
O_1^T & O_1^T & O_1^T & 0 & 0 & 0 & P & 0 \\
O_1^T & O_1^T & O_1^T & 0 & 0 & 0 & 0 & -P
\end{pmatrix}\, ,
\en
where $B$ and $Y$ are $2\times 2$ matrices (cf. with Eq.~(\ref{eq:AX})), and
$O_1$ is a $2\times 1$ matrix:
\eq
O_1=\begin{pmatrix}
0\\
0
\end{pmatrix}\, ,
\quad\quad
B=\begin{pmatrix}
E & 0\\
0 & 0
\end{pmatrix}\, ,\quad\quad
Y=\begin{pmatrix}
-\frac{1}{3}\,P-\frac{2}{3}\,S & -\frac{\sqrt{2}}{3}\,(P-S) \\
-\frac{\sqrt{2}}{3}\,(P-S) & -\frac{2}{3}\,P-\frac{1}{3}\,S
\end{pmatrix}\, ,
\en
and the quantities $K,E,P,S$ are loops with free Green functions in the non-relativistic
EFT, corresponding to the different two-particle channels:
\eq
K\bar K&:\quad& K=\int\frac{d^dk}{(2\pi)^d}\,\frac{1}{(2w_K({\bf k}))^2}\,\frac{1}{2w_K({\bf k})-P_0}\, ,
\nonumber\\[2mm]
\pi\eta&:\quad& E=\int\frac{d^dk}{(2\pi)^d}\,\frac{1}{2w_\pi({\bf k})2w_\eta({\bf k})}\,\frac{1}{w_\pi({\bf
    k})+w_\eta({\bf k})-P_0}\, ,
\nonumber\\[2mm]
\pi\pi&:\quad& P=\int\frac{d^dk}{(2\pi)^d}\,\frac{1}{(2w_\pi({\bf k}))^2}\,\frac{1}{2w_\pi({\bf k})-P_0}\, ,
\nonumber\\[2mm]
\pi \eta^s&:\quad& S=\int\frac{d^dk}{(2\pi)^d}\,\frac{1}{2w_\pi({\bf k})2w_s({\bf k})}\,\frac{1}{w_\pi({\bf
    k})+w_s({\bf k})-P_0}\, .
\en
Here, $w_s({\bf k})=(M_s^2+{\bf k}^2)^{1/2}$, where $M_s$
denotes the physical mass of the $\eta^s\doteq\bar ss$ meson, which emerges in the
partially twisted ChPT (to the lowest order, $M_s^2=2m_sB_0$, see Eq.~(\ref{eq:1-9})).

Calculating the above integrals by using the technique, described in Ref.~\cite{cusps}, we finally get
\eq
K,E,P,S=\frac{ip}{8\pi P_0}\, ,\quad\quad
p=\frac{\lambda^{1/2}(P_0^2,m_1^2,m_2^2)}{2P_0}\, .
\en
Here, $p$ stands for the relative momentum of the pair of particles in the intermediate state, $m_1,m_2$ are masses of these particles, and $\lambda(x,y,z)=x^2+y^2+z^2-2xy-2yz-2zx$ denotes the triangle function.

The potential $V$ and the $T$-matrix are also $11\times 11$ matrices. 
The $T$-matrix can be written in the
following form
\eq\label{eq:Tmatrix}
T=\left(\begin{array}{c c c c c c c c c c c} 
  c& d& \omega & \omega' & -\omega & -\omega' & b &y'& y'& y''& y''\\ 
 d& c'& \nu & \nu' & -\nu & -\nu' & b' &z'& z'& z''& z''\\ 
 \omega& \nu& f & f' & f'' & -\hat f &-\lambda & t& t'& u& u'\\ 
 \omega'& \nu'& f' & f_0 & -\hat f & f''' &-\lambda'& h& h'& r& r'\\ 
 -\omega& -\nu& f'' & -\hat f & \tilde f & \tilde f' &\lambda& -t'& -\tilde t&
-u'& -\tilde u\\ 
 -\omega'& -\nu'& -\hat f & f''' & \tilde f' & \tilde f_0 & \lambda'&-h'&
-\tilde h& -r'& -\tilde r\\ 
 b& b'& -\lambda & -\lambda' & \lambda & \lambda' & a&y& y& z& z \\
 y'& z'& t & h & -t'& -h' &y& a& y& z& z \\
 y'& z'& t' & h' & -\tilde t& -\tilde h & y& y& \tilde a & z& z \\
 y''& z''& u & r & -u'& -r' &z& z& z& q& q' \\
 y''& z''& u' & r' & -\tilde u& -\tilde r &z& z& z& q'& \tilde q \\
\end{array}\right)\, .
\en
Here, $c,d,\omega,\ldots$ denote the entries of the matrix $T_{ij}$. Some
(trivial) symmetry relations are already taken into account, for example,
$T_{36}=T_{45}=-\hat f$. Note also that the matrix $T_{ij}$ is symmetric.
On the mass shell,
the entries of the above matrix are the functions of a single Mandelstam
variable $s$ (we remind the reader
 that all partial waves except the S-wave are neglected). 
We use the name {\em physical} for the amplitudes that describe the scattering in the 
sector of only valence quarks: $T_{77}=a$ corresponds to the $K\bar K$ elastic
 scattering, $T_{11}=c$ to the $\pi\eta$ elastic scattering and
 $T_{17}=T_{71}=b$ to the $K\bar K\to\pi\eta$ transition amplitude. 
Other entries in this matrix are ``unphysical.'' For example, $y$ 
corresponds to the transition between the valence and sea quark sectors. 
Considering the quark diagrams for this process (see Fig.~\ref{fig:cd} 
and Eq.~(\ref{eq:cd}) below), one straightforwardly ensures that $y$
 corresponds to the contribution of the {\em disconnected} diagrams to 
the $K\bar K$ elastic amplitude.

There exist  more 
symmetry relations, which relate various entries in the above
matrix. A straightforward way to derive these relations in general
is to express these amplitudes in terms of the quark propagators and take into
account the fact that the valence, sea and ghost quark masses all
coincide. Below, we give few examples of such calculations.

\bigskip
{\bf Example 1:}
\medskip
Consider the quark diagrams for the transition between various $K\bar K$
states. The full 4-point Green functions of the bilinear quark operators
are given by
\eq\label{eq:cd}
G_{77}&=&t_c-t_d\, ,
\nonumber\\[2mm]
G_{88}&=&t_c-t_d\, ,
\nonumber\\[2mm]
G_{99}&=&-t_c-t_d\, ,
\nonumber\\[2mm]
G_{78}&=&G_{79}=G_{89}=G_{87}=G_{97}=G_{98}=-t_d\, ,
\en
where $t_c$ and $t_d$ denote connected and disconnected diagrams, respectively, as shown in
Fig.~\ref{fig:cd}. Different signs in different matrix elements emerge from
calculations with anti-commuting (valence, sea) and commuting (ghost) fields.
The connected diagrams are, of course, absent in the non-diagonal matrix
elements. Note that the quark propagators, used in the diagrams, are the same for
all quark species, since that masses of valence, sea and ghost quarks are the
same.

The scattering matrix elements are given by the residues of the 4-point Green
functions at the poles, corresponding to the external mesonic legs. It is 
seen that all Green functions in Eq.~(\ref{eq:cd}) are expressed only through
two quantities and, hence, there are some linear relations between them.
It can be shown (see section~\ref{sec:mixing} for the details)
that the scattering matrix elements obey exactly the same linear relations even if $\hat m\neq m_s$.
Introducing the notations $T_{77}=a$, and $T_{78}=y$, we finally arrive at the
relations 
\eq\label{eq:ex1}
T_{88}=a,\quad 
T_{99}=\tilde a=-a+2y\, ,\quad
T_{78}=T_{79}=T_{89}=T_{87}=T_{97}=T_{98}=y\, .
\en

\begin{figure}[t]
\begin{center}
\includegraphics[width=8.cm]{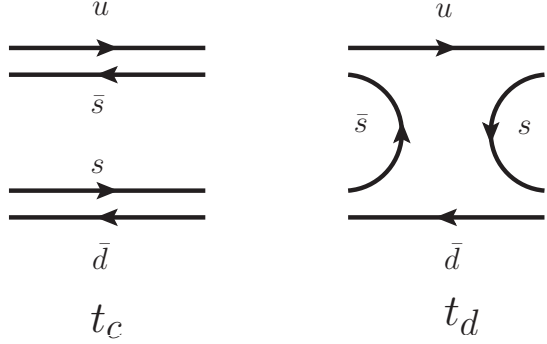}
\end{center}
\caption{Connected ($t_c$) and disconnected ($t_d$) diagrams, emerging in
  $K\bar K\to K\bar K$ scattering amplitudes with various quark species, see Eq.~(\ref{eq:cd}).}
\label{fig:cd}
\end{figure}

\bigskip
{\bf Example 2:}
\medskip

\begin{figure}[t]
\begin{center}
\includegraphics[width=10.cm]{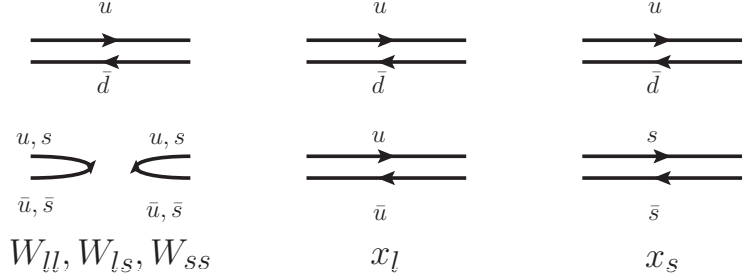}
\end{center}
\caption{
Diagrams contributing to $\pi\eta$ scattering in the valence quark
sector, see Eq.~(\ref{eq:pi-eta}). $W_{ll},W_{ls},W_{ss}$ correspond to the
diagrams with zero, one, two strange quarks in the tadpoles. $x_l$ and $x_s$ are
connected diagrams without and with strange quarks.
}
\label{fig:pi-eta}
\end{figure}

Consider
\eq\label{eq:pi-eta}
G_{33}=\frac{1}{6}\,\biggl\{[4W_{ll}-8W_{ls}+4W_{ss}]+2x_l+4x_s\biggr\}\, ,
\en
where the terms in square brackets stem from the tadpole diagrams, see
Fig.~\ref{fig:pi-eta}, and the subscripts ``$l$'' and ``$s$'' stand for 
``light'' and ``strange.''   

Carrying out similar calculations, we get
\eq
G_{34}&=&\frac{1}{3\sqrt{2}}\,\biggl\{[4W_{ll}-2W_{ls}+2W_{ss}]+2x_l-2x_s\biggr\}\,,
\nonumber\\[2mm]
G_{35}&=&-\frac{1}{6}\,\biggl\{[4W_{ll}-8W_{ls}+4W_{ss}]\biggr\}\, ,
\nonumber\\[2mm]
G_{45}&=&-\frac{1}{3\sqrt{2}}\,\biggl\{[4W_{ll}-2W_{ls}+2W_{ss}]\biggr\}\,,
\nonumber\\[2mm]
G_{44}&=&\frac{1}{3}\,\biggl\{[4W_{ll}+4W_{ls}+W_{ss}]+2x_l+x_s\biggr\}\,,
\nonumber\\[2mm]
G_{46}&=&-\frac{1}{3}\,\biggl\{[4W_{ll}+4W_{ls}+W_{ss}]\}\, ,
\nonumber\\[2mm]
G_{55}&=&\frac{1}{6}\,\biggl\{[4W_{ll}-8W_{ls}+4W_{ss}]-2x_l-4x_s\biggr\}\, ,
\nonumber\\[2mm]
G_{56}&=&\frac{1}{3\sqrt{2}}\,\biggl\{[4W_{ll}-2W_{ls}+2W_{ss}]-2x_l+2x_s\biggr\}\,,
\nonumber\\[2mm]
G_{66}&=&\frac{1}{3}\,\biggl\{[4W_{ll}+4W_{ls}+W_{ss}]-2x_l-x_s\biggr\}\,.
\en
From these relations one easily gets
\eq\label{eq:ex2}
f+\tilde f&=&-2f''\, ,
\nonumber\\[2mm]
f'+\tilde f'&=&2\hat f\, ,
\nonumber\\[2mm]
f_0+\tilde f_0&=&-2f'''\, ,
\nonumber\\[2mm]
f'-\tilde f'&=&-\sqrt{2}(f-\tilde f-f_0+\tilde f_0)\, .
\en

\bigskip

Acting in the same manner as described in the examples, we get more relations. 
The ones listed below will be used further:
\eq\label{eq:relations}
t+\tilde t&=&2t'\, ,
\nonumber\\[2mm]
y'+t-t'&=&y'+t'-\tilde t\,\,=\,\,b 
\nonumber\\[2mm]
h+\tilde h&=&2h'\, ,
\nonumber\\[2mm]
z'+h-h'&=&z'+h'-\tilde h=b'
\nonumber\\[2mm]
u+\tilde u&=&2u'\, ,
\nonumber\\[2mm]
r+\tilde r&=&2r'\, ,
\nonumber\\[2mm]
\sqrt{2}(\tilde u-u)&=&\tilde r-r\, ,
\nonumber\\[2mm]
-\sqrt{2}(\tilde h-h)&=&\tilde t-t\, ,
\nonumber\\[2mm]
\lambda&=&-t'\, ,
\nonumber\\[2mm]
\lambda'&=&-h'\, ,
\nonumber\\[2mm]
z'-\sqrt{2}y'&=&h'-\sqrt{2}t'\, ,
\nonumber\\[2mm]
\sqrt{2}d
&=&2c-2\omega+\sqrt{2}\omega'+\sqrt{2}\,(f'-\tilde f')
-(f_0-\tilde f_0)\, ,
\nonumber\\[2mm]
q+\tilde q&=&2q'\, ,
\nonumber\\[2mm]
\nu&=&\sqrt{2}\omega+\hat f+\sqrt{2}f''\, ,
\nonumber\\[2mm]
\nu'&=&\sqrt{2}\omega'-\sqrt{2}\hat f-f'''\, ,
\nonumber\\[2mm]
c'&=&c+\frac{d}{\sqrt{2}}-f'''-2f''-2\sqrt{2}\hat f-3\omega+\frac{3}{\sqrt{2}}\,\omega'\, 
\nonumber\\[2mm]
\sqrt{2}b'&=&\sqrt{2}z'+y'-b\, ,                                                
\nonumber\\[2mm]
\sqrt{2}y''-z''&=&\sqrt{2}u-r\, . 
\en
As the next step, we would like to establish, what are the 
implications of the above symmetry
relations for the potential matrix
$V_{ij}$. Recalling that in dimensional regularization the $T$-matrix
obeys the {\em  algebraic} LS equation
\eq\label{eq:LS-full}
T=V+VGT\, ,
\en
where $T$ and $G$ are given by $11\times 11$ matrices
in Eqs.~(\ref{eq:Tmatrix}) and (\ref{eq:G}), respectively. It is 
a straightforward task to solve the above matrix equation with
respect to $V$. In doing so, 
we find it useful to first perform the linear transformation of the LS equation with the matrix 
\eq
O=\left(\begin{array}{c c c c c c c c c c c }
1/2&    0  &    1/2  &      0  &    0  &    0 &    0 &    0  &   0 &
0&     0 \\
           0  &  1/2   &    0  &  1/2   &    0  &    0  &   0  &   0 &    0  &   0  &   0\\
1/4  &   0 &    - 1/4  &    0  &    1/2   &    0 &    0  &   0   &  0  &   0   &  0\\
           0  &    1/4    &   0   &    - 1/4  &    0  &    1/2  &    0  &   0  &   0 &    0  &   0\\
           1/4 &    0  &   - 1/4  &    0  &   -1/2   &    0  &   0   &  0  &   0 &    0 &    0\\
           0   &   1/4  &     0  &     -1/4   &    0  &   -1/2   &   0  &   0  &   0 &    0  &   0\\
           0  &    0 &     0  &      0   &   0  &    0 &    1  &   0  &   0  &   0 &    0\\
           0   &   0  &    0  &      0  &    0  &    0  &   0 &    1 &    0  &   0  &   0\\
           0  &    0  &    0   &     0  &    0  &    0   &  0 &    0 &    1  &   0 &    0\\
           0  &    0  &    0  &      0  &    0  &    0 &    0 &    0  &   0  &   1  &   0\\
           0  &    0  &    0     &   0   &   0   &   0 &    0 &    0  &   0 &
           0  &   1
\end{array}\right)\, .
\en
The transformed Green function is given by the matrix
\eq\label{eq:hatG}
\hat G=O^TGO=\begin{pmatrix}
B+\frac{3}{4}\,Y & -\frac{1}{4}\,Y & -\frac{1}{4}\,Y & O_1& O_1& O_1& O_1& O_1\\
-\frac{1}{4}\,Y  & -\frac{1}{4}\,Y &\frac{1}{4}\,Y   & O_1& O_1& O_1& O_1& O_1\\
-\frac{1}{4}\,Y  & \frac{1}{4}\,Y  & O_2               & O_1& O_1& O_1& O_1& O_1\\
O_1^T& O_1^T& O_1^T& K&0 &0& 0 &0\\
O_1^T& O_1^T& O_1^T & 0 &K &0& 0 &0\\
O_1^T&O_1^T& O_1^T & 0 &0 &-K& 0 &0\\
O_1^T& O_1^T& O_1^T & 0 &0 &0& P &0\\
O_1^T& O_1^T& O_1^T & 0 &0 &0& 0 &-P\\
\end{pmatrix}\, ,
\en
where
\eq
O_2=\begin{pmatrix}
0 & 0\\
0 & 0\\
\end{pmatrix}\, .
\en
Note that the entries, corresponding to the physical intermediate states
 $K\bar K$ and $\pi\eta$, appear only on the diagonal.

The transformed LS equation takes the form
\eq
\hat T=\hat V+\hat V\hat G\hat T\, ,\quad\quad
\hat T=O^{-1}T(O^T)^{-1}\, ,\quad\quad
\hat V=O^{-1}V(O^T)^{-1}\, .
\en
For the analysis of the symmetries of the potential $V$, it is convenient to further split the Green function
\eq\label{eq:splitting}
\hat G=\hat G_0+\hat G_1\, ,\quad\quad\hat G_0=\mbox{diag}\,(E,0,0,0,0,0,K,K,-K,0,0)\, .
\en
The split LS equation is:
\eq
\hat T=\hat W+\hat W \hat G_0\hat T\, ,\quad\quad
\hat W=\hat V+\hat V \hat G_1\hat W\, .
\en
It can be straightforwardly checked that in the physical matrix elements $a,b,c$
no $\pi\pi$ and $\pi \eta^s$ loops are present, namely
\eq\label{eq:noloops}
\hat V_{11}&\!\!=\!\!&\hat W_{11}\, ,\quad
\hat V_{17}=\hat W_{17}\, ,\quad
\hat V_{18}=\hat W_{18}\, ,\quad
\hat V_{19}=\hat W_{19}\, ,\quad
\hat V_{77}=\hat W_{77}\, ,\quad
\hat V_{88}=\hat W_{88}\, ,
\nonumber\\[2mm]
\hat W_{78}&\!\!=\!\!&\hat W_{79}=\hat W_{89}=\frac{1}{2}\,(\hat W_{99}+\hat V_{77})\, ,
\en
whereas the unphysical matrix elements (e.g., $y$), in general, contain such
loops:
\eq
\hat W_{78}\neq \hat V_{78}\, .
\en
In order to show this, the use of Eqs.~(\ref{eq:ex1}), (\ref{eq:ex2}) and
(\ref{eq:relations}) is crucial\footnote{Inverting $11\times 11$ matrices analytically have turned to be a very demanding task, leading to extremely lengthy expressions. What we have explicitly checked in analytic calculations 
is that the first few terms in the Born expansion of the LS equation obey the relations given in Eq.~(\ref{eq:noloops}). In addition, taking random numerical input for the $T$-matrix elements, we have checked that all symmetry relations hold {\em numerically}
for the matrix elements of the potential as well.}.

Taking into account the above relations, it is now straightforwardly seen that 
the physical matrix elements $a,b,c$ are determined from a much simpler
LS equation
\eq\label{eq:LStau}
\tau=\sigma+\sigma g\tau\, ,
\en
where $\tau,g,\sigma$ are $4\times 4$ matrices that are obtained from the
matrices $\hat T,\hat G_0,\hat W$, respectively, by deleting all rows/columns
with the indices $i,j=2,3,4,5,6,10,11$ (for these values of the indices
the matrix $\hat G_0$ has vanishing entries on the diagonal). Namely, these matrices are given by
\eq
\tau&=&\begin{pmatrix}
c & b & b & b  \\
b & a & y & y  \\
b & y & a & y  \\
b & y & y & -a+2y
\end{pmatrix}\, ,\quad
%\nonumber\\[2mm]
g=\begin{pmatrix}
E& 0 & 0 & 0 \\
0 & K & 0 & 0 \\
0 & 0 & K & 0\\
0 & 0 & 0 & -K
\end{pmatrix}\, ,\quad
%\nonumber\\[2mm]
\sigma=\begin{pmatrix}
\gamma & \beta & \beta & \beta  \\
\beta & \alpha & \delta & \delta  \\
\beta & \delta & \alpha & \delta  \\
\beta & \delta & \delta & -\alpha+2\delta
\end{pmatrix}\, ,
\nonumber\\
&&
\en
where
\eq
\alpha=\hat V_{77}\, ,\quad\beta=\hat V_{17}\, ,\quad\gamma=\hat V_{11}\, ,\quad
\delta=\hat W_{78}\, .
\en
The solution of the LS equation for the physical matrix elements gives:
\eq\label{eq:physical}
a&\!\!=\!\!&\frac{\alpha-E(\alpha\gamma-\beta^2)}{D}\, ,\quad
b=\frac{\beta}{D}\, ,\quad
c=\frac{\gamma-K(\alpha\gamma-\beta^2)}{D}\, ,
\nonumber\\[2mm]
D&\!\!=\!\!&(1-K\alpha)(1-E\gamma)-KE\beta^2\, .
\en
This solution is exactly the same as in the ``ordinary'' non-relativistic EFT 
(without sea and ghost sectors), with $\alpha,\beta,\gamma$ being the physical 
$K$-matrix elements, which we are aiming to extract from the lattice data.
Note that the physical matrix elements do not depend on the unphysical
entry $\delta$.

To summarize, in the infinite volume the solutions of the LS equation of the 
non-relativistic EFT with valence, sea and ghost sectors coincide with
those in the theory with the valence quarks only. In order to prove this
statement, it was crucial to use the symmetry relations between various physical and non-physical $T$-matrix matrix elements, which are given Eqs.~(\ref{eq:ex1}), (\ref{eq:ex2}) and (\ref{eq:relations}). This result, of course, was expected from the beginning, since these two theories are equivalent in the infinite volume.

\section{Derivation of the partially twisted L\"uscher equation}
\label{sec:luescher}

Establishing the symmetries of the potential $V$ was the most difficult part
of the problem. After this, the derivation of the partially twisted L\"uscher
equation is straightforward. The prescription, which allows one to get the 
finite-volume spectrum from the L\"uscher equation is to replace the free 
Green function $G$ by its finite-volume counterpart. Different boundary 
conditions will lead to the different modifications of $G$. On the contrary, the
potential $V$, which encodes the short-range physics, stays unaffected 
(up to exponentially suppressed contributions).

Let us consider various scenarios and see in detail, how this 
prescription works.

\bigskip
{\bf Scenario 1:}
\medskip

We impose periodic boundary conditions on the $u$-,$d$-quarks and twisted
boundary conditions on the $s$-quark:
\eq
u({\bf x}+{\bf n}L)=u({\bf x})\, ,\quad
d({\bf x}+{\bf n}L)=d({\bf x})\, ,\quad
s({\bf x}+{\bf n}L)=e^{i\mbox{\boldmath $\theta$}{\bf n}}s({\bf x})\, .
\en
These boundary conditions translate into the boundary conditions for the meson
states: the pions, etas and $\eta^s$ fields obey periodic boundary
conditions, whereas the boundary conditions for the kaons change:
\eq
K^\pm({\bf x}+{\bf n}L)=e^{\mp i\mbox{\boldmath $\theta$}{\bf n}}K^\pm({\bf x})\, ,\quad\!\!
K^0({\bf x}+{\bf n}L)=e^{-i\mbox{\boldmath $\theta$}{\bf n}}K^0({\bf x})\, ,\quad\!\!
\bar K^0({\bf x}+{\bf n}L)=e^{i\mbox{\boldmath $\theta$}{\bf n}}\bar K^0({\bf x})\, .
\en
This means that $K$ and $\bar K$ mesons {\em containing valence and ghost $s$-quarks} get additional 3-momenta $\mp\mbox{\boldmath $\theta$}/L$. The system stays in the CM frame.

The modified Green function takes the form (cf. with Eq.~(\ref{eq:hatG}))
\eq
\hat G^L=\begin{pmatrix}
B_L+\frac{3}{4}\,Y_L & -\frac{1}{4}\,Y_L & -\frac{1}{4}\,Y_L & O_1& O_1& O_1& O_1& O_1\\
-\frac{1}{4}\,Y_L  & -\frac{1}{4}\,Y_L &\frac{1}{4}\,Y_L   & O_1& O_1& O_1& O_1& O_1\\
-\frac{1}{4}\,Y_L  & \frac{1}{4}\,Y_L  & O_2               & O_1& O_1& O_1& O_1& O_1\\
O_1^T& O_1^T& O_1^T& K_L^\theta&0 &0& 0 &0\\
O_1^T& O_1^T& O_1^T & 0 &K_L &0& 0 &0\\
O_1^T&O_1^T& O_1^T & 0 &0 &-K_L^\theta& 0 &0\\
O_1^T& O_1^T& O_1^T & 0 &0 &0& P_L &0\\
O_1^T& O_1^T& O_1^T & 0 &0 &0& 0 &-P_L\\
\end{pmatrix}\, ,
\en
where the substitution rule is (cf. with Ref.~\cite{Lage-scalar})
\eq
K_L,E_L,P_L,S_L&=&\frac{1}{4\pi^{3/2} P_0L}\,Z_{00}(1;q^2)\, ,
\nonumber\\[2mm]   
K_L^\theta&=&\frac{1}{4\pi^{3/2} P_0L}\,Z_{00}^\theta(1;q^2)\, ,
\quad\quad
q=\frac{pL}{2\pi}\, .
\en
Here, $Z_{00}$ ($Z_{00}^\theta$) denotes the (twisted) L\"uscher zeta-function\eq
Z_{00}(1;q^2)&=&\frac{1}{\sqrt{4\pi}}\,\sum_{{\bf n}\in \mathbb{Z}^3}
\frac{1}{{\bf n}^2-q^2}\, ,
\nonumber\\[2mm]
Z_{00}^\theta(1;q^2)&=&\frac{1}{\sqrt{4\pi}}\,\sum_{{\bf n}\in \mathbb{Z}^3}
\frac{1}{\bigl({\bf n}+\mbox{\boldmath $\theta$}/2\pi\bigr)^2-q^2}\, .
\en
In the above equation, an ultraviolet regularization 
(e.g., the analytic regularization) is implicit.
The free Green function in a finite volume, $\hat G_L$, can be again split in analogy with Eq.~(\ref{eq:splitting}). The crucial point here is that the symmetry
of the $\hat G_{1L}$, which is the finite-volume counterpart of $\hat G_1$, remains the same. Consequently, the relations in Eq.~(\ref{eq:noloops}) still hold in a finite volume. Taking into account this fact, we can rewrite the LS 
equation~(\ref{eq:LStau}) in a finite volume:
\eq\label{eq:LStauL}
\tau_L=\sigma_L+\sigma_Lg_L\tau_L\, ,
\en
where 
\eq
g_L=\mbox{diag}\,(E_L^{},K_L^\theta,K_L^{},-K_L^\theta)\, ,
\en
and $\sigma_L$ is obtained from $\sigma$ through the replacement
$\delta\to \delta_L$. Other entries in the matrix $\sigma$, which do not contain contributions from the $\pi\pi$ and $\pi \eta^s$ loops, stay volume-independent. 

The L\"uscher equation takes the form
\eq
d_L=\mbox{det}\,(1-\sigma_Lg_L)=(1-\alpha K_L-\gamma E_L+(\alpha\gamma-\beta^2)K_L E_L)(1-(\alpha-\delta_L)K_L^\theta)^2=0\, .
\en
It is immediately seen that the determinant vanishes for those energies which
obey one of the equations
\eq\label{eq:twoeq}
0&=&1-\alpha K_L-\gamma E_L+(\alpha\gamma-\beta^2)K_L E_L\, ,
\nonumber\\[2mm]
0&=&1-(\alpha-\delta_L)K_L^\theta\, .
\en
The first equation is identical to the L\"uscher equation with {\em no twisting.} It does not depend on the non-physical entry $\delta_L$. 
The second equation
depends on the twisting angle and contains $\delta_L$. 
Since unphysical quantities appear, this equation is not very useful for the
analysis of the data.

As seen, the spectrum of the partially twisted equation contains more states
than the fully twisted one. Choosing particular source/sink operators, which do
not have an overlap with some of the states, one may project out a part of
the spectrum. For example, in our case we may consider the quark-antiquark
scalar operator ${\cal O}_s=\bar u d$, or the 4-quark operator
producing $\pi\eta$ scattering state 
${\cal O}_{\pi\eta}=(\bar u\gamma_5 d)(\bar u\gamma_5u
+\bar d\gamma_5d-2\bar s\gamma_5s)/\sqrt{6}$. It is clear that the spectrum, ``seen''
by these operators, does not depend on the twisting angle 
$\mbox{\boldmath$\theta$}$. Consequently, these operators do not overlap
with the part of the spectrum, described by the second equation in 
Eq.~(\ref{eq:twoeq}). At the level of the EFT, this is verified, e.g, from the fact
that the $\pi\eta$ scattering amplitude
\eq\label{eq:11}
(\tau_L)_{11}=\frac{\gamma-(\alpha\gamma-\beta^2)K_L}{1-\alpha K_L-\gamma E_L+(\alpha\gamma-\beta^2)K_L E_L}\, ,
\en
has poles, emerging only from the first equation in Eq.~(\ref{eq:twoeq})
(note that, say, the $K\bar K$ amplitude, which is the solution of the same
LS equation in a finite volume, contains all poles from Eq.~(\ref{eq:twoeq})).

To summarize, it is possible to derive the L\"uscher equation
with a partially twisted $s$-quark. 
The spectrum is dependent on the choice of the source/sink operators. Choosing
the operators that do not have overlap on the unphysical part of the spectrum,
it is seen that the remaining energy levels can be analyzed by using the 
L\"uscher equation {\em with no twisting at all.}
 This is not interesting, because
imposing partially twisted boundary condition does not yield new information
in this case.

\bigskip
{\bf Scenario 2:}
\medskip

Here we consider twisting of the $u$-quark, leaving
the $d$- and $s$-quarks to obey periodic boundary conditions. What changes
here is the free Green function in a finite volume.
\eq
K,E,P,S\to K_L^\theta,E_L^\theta,P_L^\theta,S_L^\theta
=\frac{1}{4\pi^{3/2}\sqrt{s}\gamma L}\,Z_{00}^{\bf d}(1;(q^*)^2)\, ,
\en
where ${\bf d}={\bf P}L/2\pi=\mbox{\boldmath$\theta$}/2\pi$,
$s=P_0^2-{\bf P}^2$, $\gamma=P_0/\sqrt{s}$, and 
\eq
q^*=\frac{p^*L}{2\pi}\, ,\quad\quad p^*=\frac{\lambda^{1/2}(s,m_1^2,m_2^2)}{2\sqrt{s}}\, .
\en
The quantity $Z_{00}^{\bf d}(1;(q^*)^2)$ denotes the L\"uscher zeta-function in the moving frame~\cite{Rummukainen}, see also Refs.~\cite{Schierholz,Hoja}:
\eq
Z_{00}^{\bf d}(1;(q^*)^2)&=&\frac{1}{\sqrt{4\pi}}\,\sum_{{\bf r}\in P_d}
\frac{1}{{\bf r}^2-(q^*)^2}\, ,
\nonumber\\[2mm]
P_d&=&\{{\bf r}=\mathbb{R}^3~|~r_\parallel=\gamma^{-1}(n_\parallel-\mu_1|{\bf d}|),
~{\bf r}_\perp={\bf n}_\perp,~{\bf n}\in\mathbb{Z}^3\}\, ,
\en
where $\mu_1=\bigl(1-(m_1^2-m_2^2)/s\bigr)/2$.

The solution of the Lippmann-Schwinger equation in a finite volume takes the following form (cf. with Eq.~(\ref{eq:physical}))
\eq
a_L&\!\!=\!\!&\frac{\alpha-E_L^\theta(\alpha\gamma-\beta^2)}{D_L^\theta}\, ,\quad
b_L=\frac{\beta}{D_L^\theta}\, ,\quad
c_L=\frac{\gamma-K_L^\theta(\alpha\gamma-\beta^2)}{D_L^\theta}\, ,
\nonumber\\[2mm]
D_L^\theta&\!\!=\!\!&(1-K_L^\theta\alpha)(1-E_L^\theta\gamma)-K_L^\theta
E_L^\theta\beta^2\, .
\en
It is seen that the spectra in case of the partial and full twisting coincide.
Both 4-quark and quark-antiquark operators couple to those eigenstates, whose
energies are described by the L\"uscher equation in the moving frame 
\eq
D_L^\theta=0\, .
\en

\bigskip
{\bf Summary:}
\medskip

Other scenarios are possible. For example, one may consider twisting $u$- and
$d$-quarks with the same angle, in order to bring two particles again in the
CM frame. We do not consider more scenarios in detail, since the pattern is
already clear from the above examples.

One observes that 
there exists the rule of thumb for the scenarios considered above.
Namely, if in a given scenario the twisted valence quarks may
annihilate (as in the scenario 1), then the corresponding partial twisting will
effectively yield no twisting. On the other hand, if the twisted valence
quarks ``go through'' all diagrams without annihilating (as in the scenario 2),
then the 
partially twisted L\"uscher equation is equivalent to the fully twisted one up
to exponentially suppressed terms. The first case is indeed easy to
understand without doing any calculations: for studying the spectrum, one
could use, for example, the quark-antiquark source and sink operators $\bar ud$,
$\bar d u$, which do not change at all, when the valence $s$-quarks are
twisted. The result in the second case looks also plausible, when one
considers quark diagrams, corresponding to the two-particle scattering
processes. However, due to technical complications, arising mainly 
from the neutral
meson mixing, certain effort is needed to elevate a plausible statement
 to a proof.

\section{Meson mixing in the neutral sector} 
\label{sec:mixing}

In the preceding sections we have derived symmetry relations for the 
elements of the scattering $T$-matrix, assuming the exact $SU(3)$-symmetric
quark content of the states corresponding to the $\eta,\eta'$ mesons:
$\eta=\eta^8\sim(u\bar u+d\bar d-2s\bar s)/\sqrt{6}$ and
$\eta'=\eta^0\sim(u\bar u+d\bar d+s\bar s)/\sqrt{3}$
in the valence, sea and ghost quark sectors (note that not all of these states
are independent due to the condition $\mbox{str}\,\Phi=0$).
This assumption holds only, if $m_s=\hat m$. At the level of the EFT, described
by the Lagrangian in Eq.~(\ref{eq:L}), the above relations hold at tree 
level and are broken by $O(p^4)$ corrections. Do our results, which rely heavily
on the symmetry relations, survive, if the  mixing is taken into
account to all orders?

The answer to this question is positive. The physical justification 
of this fact is very transparent: in the derivation of the symmetry
relations itself that involved the comparison of the quark diagrams (see 
section~\ref{sec:symmetries}), 
we have never required $m_s=\hat m$. Rather, it was assumed that 
the masses of the valence, sea and ghost quarks for each flavor coincide
(this requirement is fulfilled in our case). So, one expects that 
the results are not
affected by the breaking of the flavor $SU(3)$.

To elevate this argument to a formal level, let us consider the two-point
function of two quark bilinears in the EFT
\eq\label{eq:Dij}
D_{ij}(p^2)=i\int dxe^{ipx}\langle 0|T\chi_i(x)\chi_j(0)|0\rangle\, ,\quad\quad
i,j=1,\cdots,6\, ,
\en  
where $\chi_i=\bar\psi\Gamma_i\psi$ and the matrices $\Gamma_i$ carry all information about the spin-flavor content of the mesons. For our goals, it suffices to consider $\eta,\eta'$ mesons only (the pions and kaons do not mix).
The fermions
$\psi,\bar\psi$ belong to either valence, sea or ghost sectors.

The pole structure of $D_{ij}(p^2)$ is given by
\eq
D_{ij}(p^2)=\sum_{\alpha=1}^6\Lambda_{i\alpha}D_\alpha(p^2)
\Lambda^T_{\alpha j}+D_{ij}(p^2)^{\sf non-pole}\, ,\quad\quad D_\alpha(p^2)=\frac{c_\alpha}{M_\alpha^2-p^2}\, ,\quad c_\alpha=\pm 1,0\, , 
\en
where, at tree level, the elements of the matrix, up to a common normalization,
 $\Lambda_{i\alpha}$, can be 
read off from Eq.~(\ref{eq:phi-omega}). 
These matrix elements get modified at higher
orders in ChPT, if the flavor $SU(3)$ is broken through $\hat m\neq m_s$. 
The masses $M_\alpha^2=M_\pi^2,M_\eta^2,M_s^2$ are all equal in the
$SU(3)$ symmetry limit. At $O(p^2)$, their values can be read off from Eq.~(\ref{eq:L-omega}).

The matrix $\Lambda$, which relates the meson fields in the $SU(3)$ and 
physical bases, can be written in the following form:
\eq
\Lambda_{i\alpha}=\sum_{m=1}^6 \tilde \Lambda_{im}\stackrel{\sf 0}{\Lambda}_{m\alpha}\, ,
\en
where $\stackrel{\sf 0}{\Lambda}_{m\alpha}$ denotes the matrix at $O(p^2)$ (so far, we have
worked with this matrix), and $\tilde \Lambda_{im}$ collects all higher-order corrections.

Let us now consider the 4-point function of the quark bilinears, corresponding to the $\pi^+\eta(\eta')\to\pi^+\eta(\eta')$ scattering, see the table~1,
\eq
&&(2\pi)^4\delta^4(p_1+p_2-q_1-q_2)\,G_{ij}(p_1,p_2;q_1,q_2)
\nonumber\\[2mm]
&=&\int dx_1dx_2dy_1dy_2\,e^{ip_1x_1+ip_2x_2-iq_1y_1-iq_2y_2}
\,\langle 0|T\chi_i(x_1)\chi_{\pi^+}(x_2)\chi_j(y_1)\chi_{\pi^-}(y_2)|0\rangle .
\en
The connected piece of the 4-point function can be written as
\eq\label{eq:conn}
&&G_{ij}(p_1,p_2;q_1,q_2)^{\sf conn}
\nonumber\\[2mm]
&=&
\sum_{kl}D_{ik}(p_1^2)D_{\pi^+}(p_2^2)T_{kl}(p_1,p_2;q_1,q_2)
D_{\pi^+}(q_2^2)D_{lj}(q_1^2)
\nonumber\\[2mm]
&=&\sum_{kl}\sum_{\alpha\beta}\Lambda_{i\alpha}D_\alpha(p_1^2)\Lambda^T_{\alpha k}
D_{\pi^+}(p_2^2)T_{kl}(p_1,p_2;q_1,q_2)D_{\pi^+}(q_2^2)\Lambda_{l\beta}
D_\beta(q_1^2)\Lambda^T_{\beta j}+\cdots\, .
\en
From the above expression it is clear that the scattering amplitude in the
``physical'' basis (i.e., the basis which diagonalizes the matrix of the two-point functions), {\em on the mass shell} is given by
\eq
T^{\sf on-shell}_{\alpha\beta}(s,t)&=&\lim_{p_1^2\to M_\alpha^2,~q_1^2\to M_\beta^2,~p_2^2,q_2^2\to M_\pi^2}T_{\alpha\beta}(p_1,p_2;q_1,q_2)
\nonumber\\[2mm]
&=&\lim_{p_1^2\to M_\alpha^2,~q_1^2\to M_\beta^2,~p_2^2,q_2^2\to M_\pi^2}
\Lambda^T_{\alpha k}T_{kl}(p_1,p_2;q_1,q_2)\Lambda_{l\beta}\, ,
\en
where $s,t$ are the usual Mandelstam variables.

Now, let us consider the situation that the 4-point function of the 
quark-antiquark bilinears obeys some symmetry relations (an analogy of the
relations considered in the section~\ref{sec:symmetries})
\eq\label{eq:dG}
\sum_{ij} d_{ji}G_{ij}(p_1,p_2;q_1,q_2)=0\, ,
\en
where $d_{ij}$ are some numerical coefficients related to the structure
of the symmetry group (but not to the dynamics). Note that these are
relations that hold for {\em off-shell} momenta $p_1,p_2,q_1,q_2$.

One has to further 
distinguish between the case of the exact $SU(3)$ flavor symmetry and
broken $SU(3)$ flavor symmetry.

\bigskip
{\bf Exact \boldmath $SU(3)$ symmetry:}
\medskip

In this case $\Lambda=\stackrel{\sf 0}\Lambda$ exactly, to all orders in ChPT.
Further,  
substituting Eq.~(\ref{eq:conn}) into Eq.~(\ref{eq:dG}) and performing the
mass-shell limit, we get 
\eq\label{eq:dTab}
\sum_{\alpha\beta}k_{\beta\alpha}^{}T^{\sf on-shell}_{\alpha\beta}(s,t)=0\, ,
\quad\quad
k_{\beta\alpha}^{}=\sum_{ij}\Lambda^T_{\beta j}d_{ji}^{}\Lambda_{i\alpha}^{}
\en
We remind the reader that the disconnected
piece does not have four poles in the external momenta squared.

Next we define the on-shell amplitudes in the $SU(3)$ basis
\eq\label{eq:twobases}
T^{\sf on-shell}_{ij}(s,t)=\sum_{\alpha\beta}
\Lambda_{i\alpha}^{}T^{\sf on-shell}_{\alpha\beta}(s,t)\Lambda^T_{\beta j}\, ,
\quad\quad
\sum_{ij}d_{ji}^{}T^{\sf on-shell}_{ij}(s,t)=0\, .
\en
In other words, in case of exact $SU(3)$ symmetry, the symmetry relations
on the 4-point functions directly translate in the relations for the
on-shell amplitudes.

The LS equation in the non-relativistic EFT is derived in the basis where
the two-point function is diagonal (see section~\ref{sec:framework}).
This (matrix) equation can be written in the form
\eq
T^{\sf on-shell}_{\alpha\beta}=V_{\alpha\beta}^{}
+\sum_\gamma V_{\alpha\gamma}^{} G_\gamma^{} T^{\sf on-shell}_{\gamma\beta}\, ,
\en
where the $G_{\gamma}^{}$ are 
loops\footnote{For simplicity, we neglect here the part of the free 
Green function, which
is already diagonal, e.g., the $K\bar K$ loops. Taking them
into account does not change anything in our argumentation.}
 with the $\pi^+$ and the ``particle'' $\gamma$.
Changing now to the $SU(3)$ basis, we arrive at the equation
\eq
T^{\sf on-shell}_{ij}=V_{ij}^{}+
\sum_{nm} V_{in}^{} G_{nm}^{} T^{\sf on-shell}_{mj}\, ,
\en
where the relation between $V_{ij}^{}$ and $V_{\alpha\beta}^{}$ is the same as
between $T^{\sf on-shell}_{ij}$ and $T^{\sf on-shell}_{\alpha\beta}$, and the free
Green function in the new basis is given by
\eq
G_{ij}^{}=\sum_\gamma(\Lambda^T)^{-1}_{i\gamma}G_\gamma^{}\Lambda^{-1}_{\gamma j}\, .
\en
Our equations given in section~\ref{sec:symmetries} are exactly reproduced.
The derivation of the L\"uscher equation is straightforward. All results 
remain valid.

\bigskip
{\bf Broken \boldmath $SU(3)$ symmetry:}
\medskip

In Nature, $\hat m\neq m_s$. One may still have some exact
relations of the type given in Eq.~(\ref{eq:dG}) -- those, which do not
require $\hat m=m_s$. Examples of such relations are given in
section~\ref{sec:symmetries}. 

There are five neutral one-particle states with isospin $I=0$ 
in the physical basis (cf. with 
Eq.~(\ref{eq:basis})).
These states belong to the three different classes.
Namely, there is one state with $M_\alpha^2=M_\eta^2$, two states (one with
the wrong sign in the kinetic term) with $M_\alpha^2=M_s^2$ 
and two states (one with
the wrong sign in the kinetic term) with $M_\alpha^2=M_\pi^2$.
In Eq.~(\ref{eq:basis}), these states are described by the fields
$\omega_2$ and $\omega_3,\omega_6$ and $\omega_5,\omega_8$, respectively
(of course, the numerical values of the coefficients in this equation
are different from
the $O(p^2)$ values given in Eq.~(\ref{eq:basis})).  
We introduce a special notation for the above classes $M=\eta,\eta^s,\pi$.

Let us now consider Eq.~(\ref{eq:dG}) in the vicinity of the poles in the
momenta of the external particles. Since the masses of the particles,
belonging to the different classes, differ, if $\hat m\neq m_s$, the residues 
should vanish independently for each class. Consequently,
\eq\label{eq:M1M2}
\sum_{\mbox{state $\alpha$ in $M_1$, state $\beta$ in $M_2$}}
k_{\beta\alpha}^{}T^{\sf on-shell}_{\alpha\beta}(s,t)=0\, ,
\en
where the sum runs only over those states which belong to the classes
$M_1$ and $M_2$, respectively. For example, if $M_1=M_2=\eta$, from the above
equation we get: $k_{22} T^{\sf on-shell}_{22}=0$. If $M_1=\eta$
and $M_2=\eta^s$, we get $k_{32}(T^{\sf on-shell}_{23}+T^{\sf on-shell}_{32})
+k_{62}(T^{\sf on-shell}_{26}+T^{\sf on-shell}_{62})=0$, and so on (here, we
have used the fact that the matrix $T^{\sf on-shell}_{\alpha\beta}$ is
symmetric, as well as the matrix $k_{\alpha\beta}$).

Now, let us define
\eq
\stackrel{\sf 0}{k}_{\beta\alpha}^{}=\sum_{ij}
\stackrel{\sf 0}{\Lambda}^T_{\beta j}d_{ji}^{}
\stackrel{\sf 0}{\Lambda}_{i\alpha}^{}\, .
\en
The following crucial statement is proven in the Appendix~\ref{app:crucial}:
\begin{quote}
There is certain freedom in choosing the quantities $d_{ij}$. For example, if
we have two independent linear relations between $G_{ij}$, adding these
relations with arbitrary coefficients will yield a relation as well. Using
this freedom, one may choose the quantities $d_{ij}$ so that the following 
relation holds separately for each $M_1,M_2$
\end{quote}
\eq\label{eq:crucial}
k_{\beta\alpha}^{}=h(M_1,M_2)\stackrel{\sf 0}{k}_{\beta\alpha}^{}\, .
\en
\begin{quote}
Here, $\alpha,\beta$ label the states in the classes $M_1,M_2$, respectively,
and the number $h(M_1,M_2)$ does not depend on $\alpha$ and $\beta$.
\end{quote}

The rest of the proof is straightforward. We define the $T$-matrix in the 
$SU(3)$ basis through
\eq
T^{\sf on-shell}_{ij}(s,t)=\sum_{\alpha\beta}
\stackrel{\sf 0}{\Lambda}_{i\alpha}^{}T^{\sf on-shell}_{\alpha\beta}(s,t)
\stackrel{\sf 0}{\Lambda}^T_{\beta j}\, .
\en
We would like to stress that this is merely a {\em definition,} which is made 
for mathematical convenience only. Physically,
it does not make sense to consider a superposition of the states with different
masses in case of broken $SU(3)$ symmetry.

Using Eqs.~(\ref{eq:M1M2}) and (\ref{eq:crucial}), we easily 
derive a counterpart of Eq.~(\ref{eq:dTab}) 
\eq\label{eq:dTab0}
\sum_{\alpha\beta}\stackrel{\sf 0}{k}_{\beta\alpha}^{}
T^{\sf on-shell}_{\alpha\beta}(s,t)=0\, ,
\en
where the sum now runs over all $\alpha,\beta$ from different classes
$M_1,M_2$.

Finally, for the above definition of the $T$-matrix, one gets
\eq
\sum_{ij}d_{ji}^{}T^{\sf on-shell}_{ij}(s,t)=0\, .
\en
The free Green function in the LS equation is given by
 \eq
G_{ij}^{}=\sum_\gamma(\stackrel{\sf 0}{\Lambda}^T)^{-1}_{i\gamma}
G_\gamma^{}(\stackrel{\sf 0}{\Lambda})^{-1}_{\gamma j}\, ,
\en
and we arrive exactly at the same expressions as before. The derivation of
the L\"uscher equation is again straightforward, since only neutral mesons
with the isospin $I=0$ 
are affected by the mixing. The crucial point is that
 there is no effect of twisting for these
mesons because they are neutral. Consequently, no ambiguity arises in the
construction of the free Green function in the partially twisted case.

To summarize, the $SU(3)$ breaking affects both the free Green function and the
scattering amplitude. The matrix $\Lambda_{i\alpha}$ differs 
from its $O(p^2)$ value $\stackrel{\sf 0}{\Lambda}_{i\alpha}$. 
It is, however, possible to define the free Green function and the $T$-matrix
in the $SU(3)$ basis, still 
using the matrix  $\stackrel{\sf 0}{\Lambda}_{i\alpha}$, 
even if $\hat m\neq m_s$.
It can be now  
checked explicitly that the symmetry relations from section~\ref{sec:symmetries} 
hold for the elements of the $T$-matrix in the $SU(3)$ basis.
Consequently, our final results are unaffected by $SU(3)$ breaking. 
This was, of course, expected from the beginning, since the LS equation 
-- with the use of the above-mentioned symmetry relations -- should 
reduce to the one in the valence sector only in the infinite volume even in case of $\hat m\neq m_s$.

\section{Conclusions and outlook}
\label{sec:concl}

\begin{itemize}

\item[i)]
Using the non-relativistic EFT technique in a finite volume, we have derived 
the L\"uscher equation for the partially twisted boundary conditions for
coupled-channel $\pi\eta-K\bar K$ scattering. At an intermediate step,
the matching of the non-relativistic Lagrangian to 
partially quenched ChPT has been considered.

\item[ii)]
Our final result is remarkably simple. If in the channel with $I=I_3=1$
the light quarks are subject to twisting, the partially twisted L\"uscher
equation is equivalent to the fully twisted one, despite the presence
of annihilation diagrams. If, on the contrary, partial twisting of the
 strange quark is performed, the physically interesting part of the 
spectrum is not affected. Other scenarios are also possible and can be 
investigated by using the same methods. We think that this result is
interesting for the lattice practitioners studying the properties of
scalar mesons. We have shown that, instead of carrying out simulations at
different volumes, as required in the L\"uscher approach, one may perform
relatively cheaper partially twisted simulations.

\item[iii)]
In order to demonstrate the above result, one relies heavily on the relations 
that emerge between the various $T$-matrix elements from the valence, sea and ghost
sectors of the theory, and stem from the fact that the masses of the 
valence, sea and ghost quarks are taken equal. These relations lead to numerous 
cancellations in the 
LS equation, so that in the final equation only the physical amplitudes,
i.e., the amplitudes from the valence quark sector, are present. 
There are strong intuitive arguments, which support the above statement.
However, due to the techical complications, owing mainly to the neutral meson
mixing, a certain effort was still needed to transform these arguments
into a valid proof.

\item[iv)]
We have carried out the derivation within certain approximations. For example,
we consider only the channel with total isospin $I=1$. Moreover, all partial
waves except $l=0$ are neglected from the beginning. The partial-wave mixing
can be included later by using  standard techniques (see, e.g.,
Refs.~\cite{Leskovec,Schierholz,Liu}). 
Here, our aim was to describe
the method in the most transparent manner for one particular example, without
overloading the arguments with inessential details. 
Further, the method described
above can be used in other systems as well, for example, in the study of the
$DK$ molecules in lattice QCD (the work on this problem is in 
progress, and the results will be reported elsewhere).

\end{itemize}

{\em Acknowledgments:} 
The authors thank S. Beane, J. Bijnens, J. Gasser, 
T. L\"ahde, Ch. Liu, M. Savage, S. Sharpe
and C. Urbach for interesting discussions. 
One of us (AR) thanks the 
Institute for Nuclear Theory at the University of Washington 
for its hospitality and the Department of Energy for partial support 
during the completion of this work.
This work is partly supported by the EU
Integrated Infrastructure Initiative HadronPhysics3 Project  under Grant
Agreement no. 283286. We also acknowledge the support by the DFG (CRC 16,
``Subnuclear Structure of Matter''), by the DFG and NSFC 
(CRC 110, ``Symmetries and the Emergence
of Structure in QCD''), by
the Shota Rustaveli National Science Foundation
(Project DI/13/02) and by the Bonn-Cologne Graduate School of Physics and Astronomy.
This research is supported in part by Volkswagenstiftung
under contract no. 86260.

%\clearpage

\renewcommand{\thefigure}{\thesection.\arabic{figure}}
\renewcommand{\thetable}{\thesection.\arabic{table}}
\renewcommand{\theequation}{\thesection.\arabic{equation}}

\appendix

\setcounter{equation}{0}
\setcounter{figure}{0}
\setcounter{table}{0}

\section{Proof of Eq.~(\ref{eq:crucial})}
\label{app:crucial}

\subsection{The structure of the matrix $\Lambda_{i\alpha}$}

\begin{figure}[t]
\begin{center}
\includegraphics[width=6.cm]{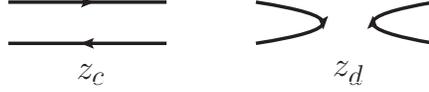}
\end{center}
\caption{Quark diagrams for the two-point function of two quark bilinears.
There are connected, $z_c$ and disconnected, $z_d$ contributions in the diagonal
matrix elements. Non-diagonal matrix elements contain only disconnected
contribution. }
\label{fig:twopointquark}
\end{figure}

The quantity $D_{ij}$ in Eq.~(\ref{eq:Dij}) is a $6\times 6$ matrix.
containing correlators of the quark bilinears
$\eta^8_{\sf vv},\eta^0_{\sf vv},\eta^8_{\sf ss},\eta^0_{\sf ss},  \eta^8_{\sf gg},
\eta^0_{\sf gg}$ where, for example,
$\eta^8_{\sf vv}=(\bar u_{\sf v}u_{\sf v}+\bar d_{\sf v}d_{\sf v}-2\bar s_{\sf v}s_{\sf v})/\sqrt{6}$ and so on.
Consider now the quark diagrams describing the two-point function of the
quark bilinears, see Fig.~\ref{fig:twopointquark}. The diagonal matrix
elements contain both connected and disconnected pieces. Keeping track of the
signs emerging in the result of (anti)commuting the fields, we get
\eq
D_{\sf vv}=D_{\sf ss}=-z_c+z_d\, ,\quad\quad
D_{\sf gg}=z_c+z_d\, .
\en
Here, all quantities are $2\times 2$ matrices.

The non-diagonal matrix elements contain only the disconnected piece:
\eq
 D_{\sf vs}=D_{\sf sg}=D_{\sf gv}=z_d\, .
\en
Taking into account the above formulae, one may conclude that the matrix 
$D_{ij}$ has, in general, the following structure (cf. with Eq.~(\ref{eq:etaprop}))\footnote{The general structure of the two-point function in the partially quenched ChPT has been discussed, e.g., in Ref.~\cite{Bijnens}}
\eq\label{eq:struct}
D_{ij}=\begin{pmatrix}
\hat A & \hat A+\hat X & \hat A+\hat X \cr
\hat A+\hat X &\hat A &\hat A+\hat X \cr
\hat A+\hat X &\hat A+\hat X &\hat A+2\hat X 
\end{pmatrix}\, ,\quad\quad
\hat A=-z_c+z_d\, ,\quad \hat X=z_c\, .
\en
Here, $\hat A,\hat X$ are $2\times 2$ matrices.

The quantity $\hat A$ in the upper left corner of the matrix $D$
is the physical propagator (it contains only valence
quarks). Consequently, it has only a pole at $p^2\to M_\eta^2$. In the
vicinity of the pole, 
\eq
\hat A_{ij}(p^2)\to \frac{\Lambda_{i\alpha}\Lambda^T_{\alpha j}}{M_\eta^2-p^2}
+\mbox{regular terms,} \quad\quad i,j=1,2\, ,\quad\quad
\mbox{state $\alpha$ in $\eta$.}
\en
Following the nomenclature of Eq.~(\ref{eq:basis}), the state in the class
$\eta$ corresponds to $\alpha=2$. Further,
choosing the proper normalization, we may write
\eq
\Lambda_{1\alpha}=\cos\tilde\theta\doteq c\, ,\quad\quad
\Lambda_{2\alpha}=\sin\tilde\theta\doteq s\, ,\quad\quad 
\mbox{state $\alpha$ in $\eta$.}
\en
We shall call $\tilde\theta$ the mixing angle. The equation~(\ref{eq:AX})
corresponds to $\tilde\theta=0$. 

Now, let us prove that the quantity $\hat X$ does not have a pole at $p^2\to
M_\eta^2$. To this end, note that the residue at the pole should be separable.
Consequently, the $2\times 2$ matrices $\hat A$, $\hat A+\hat X$, $\hat
A+2\hat X$ are all separable in the vicinity of $p^2=M_\eta^2$. Since a
separable matrix has one vanishing eigenvalue, with a orthogonal
transformation $O$ the matrix $\hat A$ can be brought to the diagonal form
$O\hat AO^T=\mbox{diag}\,(\lambda,0)$. Further, since the determinant of a
separable matrix vanishes, we have:
$\mbox{det}\,(\hat A+\hat X)=\mbox{det}\,(\hat A+2\hat X)=0$ in the vicinity
of the pole. Recalling that the matrix $\hat X$ is symmetric, 
it can be explicitly checked that this condition can be fulfilled,
if and only 
if $O\hat XO^T=\mbox{diag}\,(\lambda',0)$, i.e., $\hat X={\cal N}\,\hat A$
in the vicinity of the pole.

The value of the constant ${\cal N}$ can be fixed through the following argument.
One may change the basis $\eta^8,\eta^0$ to
$\eta^l=(\bar uu+\bar dd)/\sqrt{2}$ and $\eta^s=\bar ss$, in
valence, sea and ghost sectors. The matrix $D_{ij}$ in the new basis
has the same general structure as before. Further, it is immediately seen
that certain diagonal and non-diagonal matrix elements are equal (both contain
only disconnected contributions). For example,
\eq\label{eq:diag-nondiag}
\langle 0|T\eta^l_{\sf vv}(x)\eta^s_{\sf vv}(y)|0\rangle= 
\langle 0|T\eta^l_{\sf vv}(x)\eta^s_{\sf ss}(y)|0\rangle\, .
\en
Considering the limit $p^2\to M_\eta^2$, one may check that the above
condition is fulfilled, if and only if ${\cal N}=0$. In other words, the
matrix $\hat X$ does not have a pole at $p^2\to M_\eta^2$.

Next, we wish to demonstrate that there is no mixing, when $p^2\to M_s^2$
or $p^2\to M_\pi^2$. To this end, it is again convenient to use the basis
$\eta^l,\eta^s$ instead of $\eta^8,\eta^0$. Consider, for example, the case
$p^2\to M_s^2$. Near the pole,
\eq
D'_{ij}(p^2)\to \sum_{\mbox{state $\alpha$ in $\eta^s$}}
\frac{c_\alpha\Lambda'_{i\alpha}{\Lambda'}^T_{\alpha j}}{M_s^2-p^2}
+\mbox{regular terms}\, ,\quad\quad i,j=1,6\, ,\quad c_\alpha=\pm 1\, .
\en
Here, $D'_{ij}$ is obtained from $D_{ij}$ via the orthogonal transformation that
corresponds to the change of the basis from $\eta^8,\eta^0$ to 
$\eta^l,\eta^s$. Again following the nomenclature of Eq.~(\ref{eq:basis}), two
states in the class $\eta^s$ are $\alpha=3$ with $c_\alpha=-1$ and $\alpha=6$
with $c_\alpha=1$.

It is immediately seen that $\hat X'_{12}=\hat X'_{21}=0$ (there are no connected diagrams
for the $\eta^l-\eta^s$ transition). Also, as we shall see below, 
the pole can be contained either in $\hat X'_{11}$ or in $\hat X'_{22}$, but not in both. 
In accordance with
Eq.~(\ref{eq:AX}), we assume that the pole is contained in $\hat X'_{22}$. 
Taking into account Eq.~(\ref{eq:struct}) and the fact that $\hat A'$ does not
have a pole when $p^2\to M_s^2$, 
the following relations hold (up to an overall normalization):
\eq
&&{\Lambda'}_{23}^2-{\Lambda'}_{26}^2={\Lambda'}_{43}^2-{\Lambda'}_{46}^2=0\, ,\quad\quad
{\Lambda'}_{63}^2-{\Lambda'}_{66}^2=2\, ,
\nonumber\\[2mm]
 &&{\Lambda'}_{23}{\Lambda'}_{43}- {\Lambda'}_{26}{\Lambda'}_{46}=
{\Lambda'}_{23}{\Lambda'}_{63}- {\Lambda'}_{26}{\Lambda'}_{66}=
{\Lambda'}_{43}{\Lambda'}_{63}- {\Lambda'}_{46}{\Lambda'}_{66}=1\, .
\en
These equations still do not suffice to determine all quantities unambiguously.
To proceed further, note that $\Lambda'_{2\alpha}$ and $\Lambda'_{4\alpha}$ describe
the coupling of the $\eta^s_{\sf vv}$ and $\eta^s_{\sf ss}$ fields 
to the state $|\alpha\rangle$:
\eq
\langle 0|\bar s_{\sf v}s_{\sf v}|\alpha\rangle=\Lambda'_{2\alpha}\, ,\quad\quad
\langle 0|\bar s_{\sf s}s_{\sf s}|\alpha\rangle=\Lambda'_{4\alpha}\, .
\en
Consequently, 
\eq
\frac{1}{\sqrt{2}}\,\langle 0|(\bar s_{\sf v}s_{\sf v}\pm\bar s_{\sf s}s_{\sf s}) |\alpha\rangle=\frac{1}{\sqrt{2}}\,(\Lambda'_{2\alpha}\pm \Lambda'_{4\alpha})\, .
\en
Recall now that the operators 
$(\bar s_{\sf v}s_{\sf v}\pm\bar s_{\sf s}s_{\sf s})/\sqrt{2}$ transform differently
with respect to the {\em horizontal isospin}, corresponding to the 
$SU(2)$ rotation of the valence quarks into the sea quarks of the 
same flavor and {\em vice versa.} Horizontal isospin is a good quantum number, since the masses of the quarks of different species coincide. The physical states
$|\alpha\rangle$ should be characterized by a definite horizontal isospin. This means
that both $(\bar s_{\sf v}s_{\sf v}+\bar s_{\sf s}s_{\sf s})/\sqrt{2}$ and
$(\bar s_{\sf v}s_{\sf v}-\bar s_{\sf s}s_{\sf s})/\sqrt{2}$ can not couple to the 
same state $|\alpha\rangle$ and, consequently,
\eq
|\Lambda'_{2\alpha}|=|\Lambda'_{4\alpha}|\, ,\quad\quad \alpha=3,6\, .
\en
With this additional constraint, the above equations have the
following solution:
\eq
-{\Lambda'}_{23}={\Lambda'}_{26}=-{\Lambda'}_{43}=-{\Lambda'}_{46}=\frac{1}{\sqrt{2}}\,
  ,\quad
{\Lambda'}_{63}=-\sqrt{2}\, ,\quad {\Lambda'}_{66}=0\, .
\en
Further, since in this basis the non-diagonal matrix elements, corresponding to
the $\eta^l-\eta^s$ transition, do not have a pole, we get
\eq
{\Lambda'}_{13}={\Lambda'}_{16}={\Lambda'}_{33}={\Lambda'}_{36}={\Lambda'}_{53}={\Lambda'}_{56}=0\,
.
\en
We see that the pole can not be contained both in $\hat X'_{11}$ and 
$\hat X'_{22}$. The case of $p^2\to M_\pi^2$ is treated analogously, 
only the pole appears now in $\hat X'_{11}$.
Transforming the propagator 
back to the basis $\eta^8,\eta^0$, we finally conclude that, up
to a normalization of the quantities $D_s$ and $D_\pi$, 
 the structure of the
matrix $\hat X$ in the vicinity of the pole 
is the same as of the matrix $X$ given by Eq.~(\ref{eq:AX}).
Consequently, no mixing occurs in the matrix $\hat X$.

To summarize, the matrix $\Lambda_{i\alpha}$ has the following structure (the
index $i$ runs from 1 to 6):

\begin{itemize}
\item
The class $M=\eta$, one state
\end{itemize}
\eq\label{eq:class-eta}
\Lambda_{i,\alpha=2}\doteq n_i=\begin{pmatrix}
c \cr s \cr c \cr s \cr c\cr s 
\end{pmatrix}\, .
\en

\begin{itemize}
\item
The class $M=\eta^s$, two states
\end{itemize}
\eq\label{eq:class-S}
\Lambda_{i,\alpha=3}\doteq w^{(1)}_i=\begin{pmatrix}
-\frac{1}{\sqrt{3}}\cr \frac{1}{\sqrt{6}} \cr -\frac{1}{\sqrt{3}}\cr
\frac{1}{\sqrt{6}}\cr -\frac{2}{\sqrt{3}}\cr \frac{2}{\sqrt{6}}
\end{pmatrix}\, ,
\quad\quad
\Lambda_{i,\alpha=6}\doteq w^{(2)}_i=\begin{pmatrix}
\frac{1}{\sqrt{3}}\cr -\frac{1}{\sqrt{6}} \cr -\frac{1}{\sqrt{3}}\cr
\frac{1}{\sqrt{6}}\cr 0\cr 0
\end{pmatrix}\, .
\en

\begin{itemize}
\item
The class $M=\pi$, two states
\end{itemize}
\eq\label{eq:class-pi}
\Lambda_{i,\alpha=5}\doteq \nu^{(1)}_i=\begin{pmatrix}
-\frac{1}{\sqrt{6}}\cr -\frac{1}{\sqrt{3}} \cr \frac{1}{\sqrt{6}}\cr
\frac{1}{\sqrt{3}}\cr 0\cr 0
\end{pmatrix}\, ,
\quad\quad
\Lambda_{i,\alpha=8}\doteq \nu^{(2)}_i=\begin{pmatrix}
\frac{1}{\sqrt{6}}\cr \frac{1}{\sqrt{3}} \cr \frac{1}{\sqrt{6}}\cr
\frac{1}{\sqrt{3}}\cr \frac{2}{\sqrt{6}}\cr \frac{2}{\sqrt{3}}
\end{pmatrix}\, .
\en
The formulae for $M=\eta^s,\pi$ were read off Eq.~(\ref{eq:basis}). The common
normalization in each class is unimportant and is omitted. The
quantity $\stackrel{\sf 0}{\Lambda}_{i\alpha}$ is obtained from
$\Lambda_{i\alpha}$ by putting the mixing angle $\tilde\theta=0$, i.e.,
$c=1,s=0$.

\subsection{The linear relations between the four-point functions}

The relation given in Eq.~(\ref{eq:crucial}) holds trivially, if
$M_1=M_2=\eta$, since in this case, there is only one state. Moreover, since,
as we have found, 
the structure of $\Lambda_{i\alpha}$ is the same as of   
$\stackrel{\sf  0}{\Lambda}_{i\alpha}$, when $M=\eta^s\mbox{ or }\pi$,  
Eq.~(\ref{eq:crucial}) also holds, if both $M_1$ and $M_2$ are either
$\eta^s$ or $\pi$. What remains to be checked is the case when $M_1=\eta$
and $M_2=\eta^s\mbox{ or }\pi$.

Our strategy will be explained in few examples below. Let us start from the
identity $f+\tilde f=-2f''$, see Eq.~(\ref{eq:ex2}). The corresponding
(symmetrized) relation for the four-point functions is:
\eq
G_{33}+G_{55}+G_{53}+G_{35}=0\, .
\en
From this, one may read off the coefficients $d_{ij}$
\eq
d_{33}=d_{55}=d_{35}=d_{53}=1\, ,\quad\quad
d_{ij}=0~\mbox{otherwise}.
\en
Now, define,
\eq
k_\alpha^{(s)}=\sum_{i,j=1}^6n_iw^{(\alpha)}_jd_{ji}\, ,\quad\quad
k_\alpha^{(\pi)}=\sum_{i,j=1}^6n_i\nu^{(\alpha)}_jd_{ji}\, .
\en
Using the explicit expressions, given in Eqs.~(\ref{eq:class-eta}),
(\ref{eq:class-S}) and (\ref{eq:class-pi}), we get
\eq
k_1^{(s)}=-2c\sqrt{3}\, ,\quad\quad 
k_2^{(s)}=-\frac{2c}{\sqrt{3}}\, ,\quad\quad
k_1^{(\pi)}=\frac{2c}{\sqrt{6}}\, ,\quad\quad 
k_2^{(\pi)}=c\sqrt{6}\, .
\en
It is clear that Eq.~(\ref{eq:crucial}) is fulfilled. The factor
$h(M_1,M_2)=c$, if $M_1=\eta$ and $M_2=\eta^s\mbox{ or }\pi$.

Using the same strategy, one may verify that the Eq.~(\ref{eq:crucial}) holds
also for the following linear relations (cf. with section~\ref{sec:symmetries}):
\eq
f_0+\tilde f_0&=&-2f'''\, ,
\nonumber\\[2mm]
f'+\tilde f'&=&2\hat f\, ,
\nonumber\\[2mm]
f'-\tilde f'&=&-\sqrt{2}(f-\tilde f-f_0+\tilde f_0)\, .
\en
The relation
\eq\label{eq:relation}
\sqrt{2}d=2c-2\omega+\sqrt{2}\omega'+\sqrt{2}(f'-\tilde f')-(f_0-\tilde f_0)
\en
is more complicated. Using the identity $G_{36}-G_{45}+G_{63}-G_{54}=0$,
we may rewrite Eq.~(\ref{eq:relation}) the following form:
\eq
 &&\frac{1}{\sqrt{2}}\,(G_{12}+G_{21})-2G_{11}+(G_{13}+G_{31})
-\frac{1}{\sqrt{2}}\,(G_{14}+G_{41})
\nonumber\\[2mm]
&-&\frac{1}{\sqrt{2}}\,(G_{34}+G_{43}-G_{56}+G_{65})
+a(G_{36}-G_{45}+G_{63}-G_{54})=0\, ,
\en
where $a$ is arbitrary.
Reading off the coefficients $d_{ij}$ from the above equation, one may verify
by direct calculations that Eq.~(\ref{eq:crucial}) holds, if the choice
$a=\sqrt{2}$ is made.

We have further checked that the remaining identities
\eq
c'&=&c+\frac{d}{\sqrt{2}}-f'''-2f''-2\sqrt{2}\hat f-3\omega
+\frac{3}{\sqrt{2}}\,\omega'\, ,
\nonumber\\[2mm]
\nu&=&\sqrt{2}\omega+\hat f+\sqrt{2}f''\, ,
\nonumber\\[2mm]
\nu'&=&\sqrt{2}\omega'-\sqrt{2}\hat f-f'''\, ,
\en
can be treated in a similar fashion. Adding the term
$a(G_{36}-G_{45}+G_{63}-G_{54})$ to the pertinent linear
relations for the four-point functions, it is shown that the constant $a$ can be
always adjusted so that the Eq.~(\ref{eq:crucial}) is fulfilled. 

In section~\ref{sec:symmetries} more linear relations are displayed, 
which correspond to the transitions involving the states not affected
by mixing. It is a straightforward task to verify that the same arguments
can be applied in this case as well.


\newpage

\begin{thebibliography}{99}


\bibitem{Kunihiro}
  T.~Kunihiro {\it et al.}  [SCALAR Collaboration],
  %``Scalar mesons in lattice QCD,''
  Phys.\ Rev.\ D {\bf 70} (2004) 034504
  [hep-ph/0310312].
  %%CITATION = HEP-PH/0310312;%%


\bibitem{Hart}
  C.~McNeile {\it et al.}  [UKQCD Collaboration],
  %``Properties of light scalar mesons from lattice QCD,''
  Phys.\ Rev.\ D {\bf 74} (2006) 014508
  [hep-lat/0604009];
  %%CITATION = HEP-LAT/0604009;%%


  A.~Hart {\it et al.}  [UKQCD Collaboration],
  %``A Lattice study of the masses of singlet 0++ mesons,''
  Phys.\ Rev.\ D {\bf 74} (2006) 114504
  [hep-lat/0608026].
  %%CITATION = HEP-LAT/0608026;%%


\bibitem{Prelovsek}  

S.~Prelovsek and D.~Mohler,
  %``A Lattice study of light scalar tetraquarks,''
  Phys.\ Rev.\ D {\bf 79} (2009) 014503
  [arXiv:0810.1759 [hep-lat]];
  %%CITATION = ARXIV:0810.1759;%%

  S.~Prelovsek, T.~Draper, C.~B.~Lang, M.~Limmer, K.~-F.~Liu, N.~Mathur and D.~Mohler,
  %``Lattice study of light scalar tetraquarks with I=0,2,1/2,3/2: Are \sigma and \kappa tetraquarks?,''
  Phys.\ Rev.\ D {\bf 82} (2010) 094507
  [arXiv:1005.0948 [hep-lat]].
  %%CITATION = ARXIV:1005.0948;%%


\bibitem{Urbach}
  C.~Alexandrou, J.~O.~Daldrop, M.~Dalla Brida, M.~Gravina, L.~Scorzato, C.~Urbach and M.~Wagner,
  %``Lattice investigation of the scalar mesons a_0(980) and \kappa\ using four-quark operators,''
  JHEP {\bf 1304} (2013) 137
  [arXiv:1212.1418 [hep-lat]];
  %%CITATION = ARXIV:1212.1418;%%

  M.~Wagner, C.~Alexandrou, J.~O.~Daldrop, M.~D.~Brida, M.~Gravina, L.~Scorzato, C.~Urbach and C.~Wiese,
  %``Scalar mesons and tetraquarks from twisted mass lattice QCD,''
  arXiv:1302.3389 [hep-lat].
  %%CITATION = ARXIV:1302.3389;%%
\bibitem{Luescher-torus}
  M.~L\"uscher,
  %``Two particle states on a torus and their relation to the scattering matrix,''
  Nucl.\ Phys.\ B {\bf 354} (1991) 531.
  %%CITATION = NUPHA,B354,531;%%
  %442 citations counted in INSPIRE as of 14 Jun 2013

\bibitem{Lage-scalar}
  V.~Bernard, M.~Lage, U.-G.~Mei{\ss}ner and A.~Rusetsky,
  %``Scalar mesons in a finite volume,''
  JHEP {\bf 1101} (2011) 019
  [arXiv:1010.6018 [hep-lat]].
  %%CITATION = ARXIV:1010.6018;%%

\bibitem{Oset-scalar1}
  M.~D\"oring, U.-G.~Mei{\ss}ner, E.~Oset and A.~Rusetsky,
  %``Unitarized Chiral Perturbation Theory in a finite volume: Scalar meson sector,''
  Eur.\ Phys.\ J.\ A {\bf 47} (2011) 139
  [arXiv:1107.3988 [hep-lat]].
  %%CITATION = ARXIV:1107.3988;%%


\bibitem{Oset-scalar2}
  M.~D\"oring, U.-G.~Mei{\ss}ner, E.~Oset and A.~Rusetsky,
  %``Scalar mesons moving in a finite volume and the role of partial wave mixing,''
  Eur.\ Phys.\ J.\ A {\bf 48} (2012) 114
  [arXiv:1205.4838 [hep-lat]].
  %%CITATION = ARXIV:1205.4838;%%


\bibitem{Jaffe:1976ig}
  R.~L.~Jaffe,
  %``Multi-Quark Hadrons. 1. The Phenomenology Of (2 Quark 2 Anti-Quark)
  %Mesons,''
  Phys.\ Rev.\  D {\bf 15} (1977) 267.
  %%CITATION = PHRVA,D15,267;%%

\bibitem{Black:1998wt}
  D.~Black {\it et al},
% A.~H.~Fariborz, F.~Sannino and J.~Schech\-ter,
  %``Putative Light Scalar Nonet,''
  Phys.\ Rev.\  D {\bf 59} (1999) 074026
  [arXiv:hep-ph/9808415].
  %%CITATION = PHRVA,D59,074026;%%

\bibitem{Achasov:2002ir}
  N.~N.~Achasov and A.~V.~Kiselev,
  %``The new analysis of the KLOE data on the phi->eta pi0 gamma decay,''
  Phys.\ Rev.\  D {\bf 68} (2003) 014006
  [arXiv:hep-ph/0212153].
  %%CITATION = PHRVA,D68,014006;%%


\bibitem{Pelaez:2004xp}
  J.~R.~Pelaez,
  %``Light scalars as tetraquarks or two-meson states from large Nc and
  %unitarized Chiral Perturbation Theory,''
  Mod.\ Phys.\ Lett.\  A {\bf 19} (2004) 2879
  [arXiv:hep-ph/0411107].
  %%CITATION = MPLAE,A19,2879;%%


\bibitem{Weinstein:1982gc}
  J.~D.~Weinstein and N.~Isgur,
  %``Do Multi-Quark Hadrons Exist?,''
  Phys.\ Rev.\ Lett.\  {\bf 48} (1982) 659.
  %%CITATION = PRLTA,48,659;%%



\bibitem{Oller:1997ng}
  J.~A.~Oller and E.~Oset,
  %``Chiral Symmetry Amplitudes in the S-Wave Isoscalar and Isovector Channels
  %and the \sigma, f_0(980), a_0(980) Scalar Mesons,''
  Nucl.\ Phys.\  A {\bf 620} (1997) 438
  [Erratum-ibid.\  A {\bf 652} (1999) 407]
  [arXiv:hep-ph/9702314];
  %%CITATION = NUPHA,A620,438;%%


\bibitem{oop}
  J.~A.~Oller, E.~Oset and J.~R.~Pelaez,
  %``Non-perturbative Approach to effective chiral Lagrangians and Meson
  %Interactions,''
  Phys.\ Rev.\ Lett.\  {\bf 80} (1998) 3452
  [arXiv:hep-ph/9803242];
  %%CITATION = PRLTA,80,3452;%%


  J.~A.~Oller, E.~Oset and J.~R.~Pelaez,
  %``Meson-Meson interaction in a non-perturbative chiral approach,''
  Phys.\ Rev.\  D {\bf 59} (1999) 074001
  [Erratum-ibid.\  D {\bf 60} (1999\ ERRAT,D75,099903.2007) 099906]
  [arXiv:hep-ph/9804209].
  %%CITATION = PHRVA,D59,074001;%%


\bibitem{Oller:1998zr}
  J.~A.~Oller and E.~Oset,
  %``N/D Description of Two Meson Amplitudes and Chiral Symmetry,''
  Phys.\ Rev.\  D {\bf 60} (1999) 074023
  [arXiv:hep-ph/9809337];
  %%CITATION = PHRVA,D60,074023;%%

  J.~A.~Oller, E.~Oset and J.~R.~Pelaez,
  %``Meson-Meson interaction in a non-perturbative chiral approach,''
  Phys.\ Rev.\  D {\bf 59} (1999) 074001
  [Erratum-ibid.\  D {\bf 60} (1999\ ERRAT,D75,099903.2007) 099906]
  [arXiv:hep-ph/9804209].
  %%CITATION = PHRVA,D59,074001;%%



\bibitem{sumrules}
  S.~Peris, M.~Perrottet and E.~de Rafael,
  %``Matching long and short distances in large-N(c) {QCD},''
  JHEP {\bf 9805} (1998) 011
  [arXiv:hep-ph/9805442];
  %%CITATION = JHEPA,9805,011;%%


  V.~Elias, A.~H.~Fariborz, F.~Shi and T.~G.~Steele,
  %``{QCD} sum rule consistency of lowest-lying q anti-q scalar resonances,''
  Nucl.\ Phys.\  A {\bf 633} (1998) 279
  [arXiv:hep-ph/9801415].
  %%CITATION = NUPHA,A633,279;%%



\bibitem{Janssen:1994wn}
  G.~Janssen, B.~C.~Pearce, K.~Holinde and J.~Speth,
  %``On the structure of the scalar mesons $f_0(975)$ and $a_0(980)$,''
  Phys.\ Rev.\  D {\bf 52} (1995) 2690
  [arXiv:nucl-th/9411021].
  %%CITATION = PHRVA,D52,2690;%%

\bibitem{Weinberg:1963zz}
  S.~Weinberg,
Phys.\ Rev.\  {\bf 130} (1963) 776;
  Phys.\ Rev.\  {\bf 131} (1963) 440;
Phys.\ Rev.\  {\bf 137} (1965) B672.
  %%CITATION = PHRVA,131,440;%%

\bibitem{Morgan:1992ge}
  D.~Morgan,
  %``Pole Counting And Resonance Classification,''
  Nucl.\ Phys.\  A {\bf 543} (1992) 632.
  %%CITATION = NUPHA,A543,632;%%


\bibitem{Tornqvist:1994ji}
  N.~A.~T\"ornqvist,
  %``How To Parametrize An S Wave Resonance And How To Identify Two Hadron
  %Composites,''
  Phys.\ Rev.\  D {\bf 51} (1995) 5312
  [arXiv:hep-ph/9403234].
  %%CITATION = PHRVA,D51,5312;%%



\bibitem{Morgan:1993td}

  D.~Morgan and M.~R.~Pennington,
  %``f0 (S*): Molecule or quark state?,''
  Phys.\ Lett.\  B {\bf 258} (1991) 444
  [Erratum-ibid.\  B {\bf 269} (1991) 477];
  %%CITATION = PHLTA,B258,444;%%
  %``New data on the K anti-K threshold region and the nature of the f0 (S*),''
  Phys.\ Rev.\  D {\bf 48} (1993) 1185.
  %%CITATION = PHRVA,D48,1185;%%


\bibitem{Baru:2003qq}
  V.~Baru,
   J.~Haidenbauer, C.~Hanhart, Yu.~Kalashnikova and A.~E.~Kudryavtsev,
  %``Evidence that the a0(980) and f0(980) are not elementary particles,''
  Phys.\ Lett.\  B {\bf 586} (2004) 53
  [arXiv:hep-ph/0308129].
  %%CITATION = PHLTA,B586,53;%%



\bibitem{Baru:2004xg}
  V.~Baru, J.~Haidenbauer, C.~Hanhart, A.~E.~Kudryavtsev and U.-G.~Mei{\ss}ner,
  %``Flatte-like distributions and the a0(980) / f0(980) mesons,''
  Eur.\ Phys.\ J.\  A {\bf 23} (2005) 523
  [arXiv:nucl-th/0410099].
  %%CITATION = EPHJA,A23,523;%%


\bibitem{Hanhart:2006nr}
  C.~Hanhart,
  %``Towards an understanding of the light scalar mesons,''
  Eur.\ Phys.\ J.\  A {\bf 31} (2007) 543
  [arXiv:hep-ph/0609136].
  %%CITATION = EPHJA,A31,543;%%


\bibitem{twisted}
  P.~F.~Bedaque,
  %``Aharonov-Bohm effect and nucleon nucleon phase shifts on the lattice,''
  Phys.\ Lett.\  B {\bf 593} (2004) 82
  [arXiv:nucl-th/0402051];
  %%CITATION = PHLTA,B593,82;%%


  G.~M.~de Divitiis, R.~Petronzio and N.~Tantalo,
  %``On the discretization of physical momenta in lattice QCD,''
  Phys.\ Lett.\  B {\bf 595} (2004) 408
  [arXiv:hep-lat/0405002];
  %%CITATION = PHLTA,B595,408;%%


  G.~M.~de Divitiis and N.~Tantalo,
  %``Non leptonic two-body decay amplitudes from finite volume calculations,''
  arXiv:hep-lat/0409154.
  %%CITATION = HEP-LAT/0409154;%%


\bibitem{Sachrajda}

  C.~T.~Sachrajda and G.~Villadoro,
  %``Twisted boundary conditions in lattice simulations,''
  Phys.\ Lett.\  B {\bf 609} (2005) 73
  [arXiv:hep-lat/0411033].
  %%CITATION = PHLTA,B609,73;%%

\bibitem{Chen}
  P.~F.~Bedaque and J.-W.~Chen,
  %``Twisted valence quarks and hadron interactions on the lattice,''
  Phys.\ Lett.\ B {\bf 616} (2005) 208
  [hep-lat/0412023].
  %%CITATION = HEP-LAT/0412023;%%
 


\bibitem{Beane}
  S.~R.~Beane, P.~F.~Bedaque, A.~Parreno and M.~J.~Savage,
  %``Exploring Hyperons And Hypernuclei With Lattice QCD,''
  Nucl.\ Phys.\ A {\bf 747} (2005) 55
  [arXiv:nucl-th/0311027].
  %%CITATION = NUCL-TH 0311027;%%

\bibitem{Lage-piN}
  V.~Bernard, M.~Lage, U.-G.~Mei{\ss}ner and A.~Rusetsky,
  %``Resonance properties from the finite-volume energy spectrum,''
  JHEP {\bf 0808} (2008) 024
  [arXiv:0806.4495 [hep-lat]].
  %%CITATION = ARXIV:0806.4495;%%
  %52 citations counted in INSPIRE as of 15 Jun 2013

\bibitem{Hoja}
  V.~Bernard, D.~Hoja, U.-G.~Mei{\ss}ner and A.~Rusetsky,
  %``Matrix elements of unstable states,''
  JHEP {\bf 1209} (2012) 023
  [arXiv:1205.4642 [hep-lat]].
  %%CITATION = ARXIV:1205.4642;%%
  %7 citations counted in INSPIRE as of 15 Jun 2013

\bibitem{Sharpe}
  S.~R.~Sharpe and N.~Shoresh,
  %``Partially quenched chiral perturbation theory without Phi0,''
  Phys.\ Rev.\ D {\bf 64} (2001) 114510
  [hep-lat/0108003].
  %%CITATION = HEP-LAT/0108003;%%
  %143 citations counted in INSPIRE as of 15 Jun 2013

\bibitem{physrep}
  J.~Gasser, V.~E.~Lyubovitskij and A.~Rusetsky,
  %``Hadronic atoms in QCD + QED,''
  Phys.\ Rept.\  {\bf 456} (2008) 167\\{}
  [arXiv:0711.3522 [hep-ph]].
  %%CITATION = ARXIV:0711.3522;%%
  %55 citations counted in INSPIRE as of 17 Jun 2013


\bibitem{cusps}
  J.~Gasser, B.~Kubis and A.~Rusetsky,
  %``Cusps in K --> 3pi decays: a theoretical framework,''
  Nucl.\ Phys.\ B {\bf 850} (2011) 96
  [arXiv:1103.4273 [hep-ph]].
  %%CITATION = ARXIV:1103.4273;%%
  %17 citations counted in INSPIRE as of 17 Jun 2013

\bibitem{Rummukainen}
  K.~Rummukainen and S.~A.~Gottlieb,
  %``Resonance scattering phase shifts on a nonrest frame lattice,''
  Nucl.\ Phys.\ B {\bf 450} (1995) 397
  [hep-lat/9503028].
  %%CITATION = HEP-LAT/9503028;%%

\bibitem{Schierholz}
  M.~G\"ockeler, R.~Horsley, M.~Lage, U.-G.~Mei{\ss}ner, P.~E.~L.~Rakow, A.~Rusetsky, G.~Schierholz and J.~M.~Zanotti,
  %``Scattering phases for meson and baryon resonances on general moving-frame lattices,''
  Phys.\ Rev.\ D {\bf 86} (2012) 094513
  [arXiv:1206.4141 [hep-lat]].
  %%CITATION = ARXIV:1206.4141;%%

\bibitem{Leskovec}
  L.~Leskovec and S.~Prelovsek,
  %``Scattering phase shifts for two particles of different mass and non-zero total momentum in lattice QCD,''
  Phys.\ Rev.\ D {\bf 85} (2012) 114507
  [arXiv:1202.2145 [hep-lat]].
  %%CITATION = ARXIV:1202.2145;%%

\bibitem{Liu}
  N.~Li and C.~Liu,
  %``Generalized L\'uscher Formula in Multi-channel Baryon-Meson Scattering,''
  Phys.\ Rev.\ D {\bf 87} (2013) 014502
  [arXiv:1209.2201 [hep-lat]].
  %%CITATION = ARXIV:1209.2201;%%
  %3 citations counted in INSPIRE as of 22 Jul 2013





\bibitem{Bijnens}
  J.~Bijnens and N.~Danielsson,
  %``The Eta mass and NNLO three-flavor partially quenched chiral perturbation theory,''
  Phys.\ Rev.\ D {\bf 74} (2006) 054503
  [hep-lat/0606017].
  %%CITATION = HEP-LAT/0606017;%%



\end{thebibliography}
\end{document}